\documentclass[a4paper,fleqn,usenatbib]{mnras}

\usepackage{mathptmx}
\usepackage{geometry}

\usepackage[T1]{fontenc}
\usepackage{ae,aecompl}

\usepackage{graphicx}	% Including figure files
\usepackage{amsmath}	% Advanced maths commands
\usepackage{amssymb}	% Extra maths symbols
\usepackage{bm}
\usepackage{newtxtext,newtxmath}
\usepackage{xcolor}
\usepackage{xspace}
\usepackage[normalem]{ulem}

\newcommand{\vektor}[1]{\mathbfit{#1}}
\newcommand{\tensor}[1]{\bm{\mathsf{#1}}}
\newcommand{\vnabla}{\bm{\nabla}}
\newcommand{\bcdot}{\bm{\cdot}}
\newcommand{\dd}{\mathrm{d}}
\newcommand{\slope}{q}
\newcommand{\cri}{{\mathrm{cr},i}}
\newcommand{\cra}{{\mathrm{cr}}}
\newcommand{\pp}{\partial}
\newcommand{\gpcc}{\mathrm{g\,cm^{-3}}}
\newcommand{\GeVc}{\mathrm{GeV}~c^{-1}}
\newcommand{\TeVc}{\mathrm{TeV}~c^{-1}}

\newcommand{\rkl}[1]{\left(#1\right)}
\newcommand{\ekl}[1]{\left[#1\right]}

\newcommand{\skl}[1]{\left\langle#1\right\rangle}

\newcommand{\greyadv}{\texttt{grey-adv}\xspace}
\newcommand{\greydiff}{\texttt{grey-diff-1e28}\xspace}
\newcommand{\greydiffhi}{\texttt{grey-diff-4e28}\xspace}
\newcommand{\spec}{\texttt{spectral}\xspace}

\newcommand{\B}{\mathcal{B}}
\newcommand{\p}{\rmn{p}}

\newcommand{\bvel}{\ensuremath{\bm{\varv}}}

\defcitealias{GirichidisEtAl2020}{Paper I}

\title[Spectral CR-MHD - II]{Spectrally resolved cosmic rays - II. Momentum-dependent cosmic ray diffusion drives powerful galactic winds}

\author[Philipp Girichidis et al.]{
  Philipp~Girichidis$^{1,2}$\thanks{E-mail: philipp@girichidis.com},
  Christoph~Pfrommer$^{1}$, R\"{u}diger Pakmor$^3$, Volker Springel$^3$
\\
$^{1}$Leibniz-Institut f\"{u}r Astrophysik (AIP), An der Sternwarte 16, 14482 Potsdam, Germany\\
$^{2}$Universit\"{a}t Heidelberg, Zentrum f\"{u}r Astronomie, Institut f\"{u}r Theoretische Astrophysik, Albert-Ueberle-Str. 2, D-69120 Heidelberg, Germany\\
$^{3}$Max-Planck-Institut f\"{u}r Astrophysik, Karl-Schwarzschild-Str. 1, 85741 Garching, Germany\\
}

\date{Accepted XXX. Received YYY; in original form ZZZ}

\pubyear{2020}

\begin{document}
\label{firstpage}
\pagerange{\pageref{firstpage}--\pageref{lastpage}}
\maketitle

\begin{abstract}
Cosmic ray (CR) feedback has been identified as a critical process in galaxy formation. Most previous simulations have integrated out the energy-dependence of the CR distribution, despite its large extent over more than twelve decades in particle energy. To improve upon this simplification, we present the implementation and first application of spectrally resolved CRs which are coupled to the magneto-hydrodynamics in simulations of galaxy formation. The spectral model for the CRs enables more accurate CR cooling and allows for an energy-dependent spatial diffusion, for which we introduce a new stable numerical algorithm that proves essential in highly dynamical systems. We perform galaxy formation simulations with this new model and compare the results to a grey CR approach with a simplified diffusive transport and effective cooling that assumes steady-state spectra. We find that the galaxies with spectrally resolved CRs differ in morphology, star formation rate, and strength and structure of the outflows. The first outflow front is driven by CRs with average momenta of $\sim200-600\,\mathrm{GeV}~c^{-1}$. The subsequent formation of outflows, which reach mass loading factors of order unity, are primarily launched by CRs of progressively smaller average momenta of $\sim8-15\,\mathrm{GeV}~c^{-1}$. The CR spectra in the galactic centre quickly approach a steady state, with small temporal variations. In the outer disc and outflow regions, the spectral shape approaches steady state only after $\sim2\,\mathrm{Gyr}$ of evolution. Furthermore, the shapes of the approximate steady state spectra differ for individual regions of the galaxy, which highlights the importance of actively including the full CR spectrum.
\end{abstract}

\begin{keywords}
cosmic rays, (magnetohydrodynamics) MHD, galaxies: formation, galaxies: evolution, methods: numerical, diffusion
\end{keywords}

%%%%%%%%%%%%%%%%%%%%%%%%%%%%%%%%%%%%%%%%%%%%%%%%%%

%%%%%%%%%%%%%%%%% BODY OF PAPER %%%%%%%%%%%%%%%%%%

\section{Introduction}

CRs are an important non-thermal energy component in galaxies \citep{StrongMoskalenkoPtuskin2007, GrenierBlackStrong2015,Zweibel2017} with an energy density that is comparable to that of the thermal, magnetic and turbulent energy reservoirs \citep{Cox2005, NaabOstriker2017}. Particularly, their transport mechanisms and their interactions with electromagnetic waves couple the collisionless CR population dynamically to the thermal plasma \citep{Zweibel2013,Zweibel2017} and cause a large variety of dynamical effects in and around galaxies from the regulation of star formation in the densest regions of the galaxy \citep[e.g.][]{PadovaniEtAl2020} to large-scale galactic outflows \citep[e.g.][]{VeilleuxEtAl2020} to providing a significant fraction of the pressure in the circum-galactic medium \citep{Salem2016,BuckEtAl2020,Ji2020}. The individual effects are linked to different energy ranges of CRs. Low-energy CRs -- mostly non-relativistic CR protons with momenta $\lesssim0.1\,\GeVc$ -- have an increased cross section with the thermal gas, which causes them to cool efficiently, transfer their energy to the gas and ionize neutral or even molecular gas \citep{IvlevEtAl2018,PhanMorlinoGabici2018}. This causes a temperature floor in the dense gas, which regulates the fragmentation of gas and the resulting formation of stars \citep{PadovaniEtAl2020}. The main dynamical driver in galaxies are CRs in the GeV range, which cool less efficiently and are able to transport a significant amount of energy to the interstellar medium (ISM) and drive galactic outflows \citep{Ipavich1975, Breitschwerdt1991, ZirakashviliEtAl1996, PtuskinEtAl1997}. High-energy CRs (with momenta $\gtrsim\TeVc$) are overall subdominant in integrated energy, but are an important source of non-thermal radiative signatures \citep{KoteraOlinto2011,WerhahnEtAl2021b,WerhahnEtAl2021c}.

Numerical simulations have investigated the dynamical impact of CRs in a large variety of setups and conditions. Early models included the non-thermal energy in a simplified advection-diffusion approximation and found CR-driven instabilities like the CR counterpart to the Parker instability \citep{HanaszLesch2003}. Applications of the advection-diffusion approximations in entire galaxies revealed their ability to drive galactic outflows, which has been suggested earlier by one-dimensional models \citep{Breitschwerdt1991,Everett2008,DorfiBreitschwerdt2012, RecchiaBlasiMorlino2016}. The variety of galactic models is large and ranges from non-magnetic simulations with isotropic CR diffusion \citep[e.g.,][]{JubelgasEtAl2008,SalemBryan2014, BoothEtAl2013, SemenovKravtsovCaprioli2021}, CR streaming \citep[e.g.,][]{UhligEtAl2012,WienerEtAl2017} and magneto-hydrodynamic (MHD) models with anisotropic CR diffusion along the magnetic field lines \citep{YangEtAl2012,HanaszEtAl2013,PakmorEtAl2016,PfrommerEtAl2017,PeschkenEtAl2021}. The effects of the streaming instability of CRs in galactic models has also been considered \citep{RuszkowskiYangZweibel2017, ChanEtAl2019, DashyanDubois2020, HopkinsEtAl2021}. Besides full galactic models CRs have also been included in ISM models, which explicitly resolve the multi-phase structure and solve for chemical processes in the supernova (SN)-driven ISM \citep{GirichidisEtAl2016a, SimpsonEtAl2016, GirichidisEtAl2018a, FarberEtAl2018, CommerconMarcowithDubois2019, ArmillottaOstrikerJiang2021}. The exact driving mechanism, strength, and halo-mass dependence of the outflows \citep{JacobEtAl2018} is still a matter of debate and depends significantly on the other feedback processes in the ISM \citep[e.g.][]{RathjenEtAl2021} and on the models for CR transport that go beyond the one-moment advection-diffusion approximation \citep[e.g.][]{JiangOh2018,ThomasPfrommer2019,ThomasPfrommer2021,ThomasEtAl2021}. Those more accurate descriptions of CR physics are supported by MeerKAT radio obervations of the radio brightness distribution along faint non-thermal filaments pervading the central molecular zone in our Galaxy \citep{ThomasEtAl2020}.

However, all previous models only include a grey approach for CR transport, i.e. they follow the momentum-integrated particle distribution function and model CRs as an effective fluid with an effective transport coefficient and effective cooling properties, independent of individual CR momenta. The basis for this effective model is the assumption of a steady state spectrum. However, including the spectral CR description in the post-processing of effective grey models suggests that for only a fraction of the galaxy the assumption of a steady state is appropriate. In particular for the highly dynamical outflow region the steady state assumption breaks down \citep{WerhahnEtAl2021a}. In order to accurately solve the dynamical effects as well as CR cooling and ionization, we need to simulate galaxy formation and evolution with spectrally resolved CRs. In \citeauthor{GirichidisEtAl2020} (\citeyear{GirichidisEtAl2020}, hereafter Paper I), we have presented an efficient numerical method that allows to solve for a dynamically evolving CR spectrum in every computational cell and accurately couples the locally varying CR pressure to the gas dynamics.

In this study we implement the spectral solver from \citetalias{GirichidisEtAl2020} into the Arepo moving mesh code \citep{Springel2010,WeinbergerSpringelPakmor2020}. We improve on the energy dependent diffusion model in order to increase the stability of the method in highly dynamical regions and discuss the changes in the adiabatic index. We then apply the spectrally resolved models to simulations of an intermediate-mass galaxy with a halo mass of $10^{11}~\rmn{M}_\odot$ and compare our newly developed spectral CR model to previous grey approaches. For the first time, we will demonstate  that spectrally resolved CRs have a significant dynamical impact that differs from the grey approaches in terms of star formation rate, the morphology of the galaxy as well as the onset and structure of outflows.

The outline of the paper is as follows. In Section~\ref{sec:numerics}, we introduce our new spectral CR model, describe the new spatial CR diffusion algorithm and introduce the generalized adiabatic index that is a necessary consequence of our spectrally resolved CR distribution. In Section~\ref{sec:code}, we outline our simulation setup and describe the models presented in this work. In Section~\ref{sec:global}, we compare the global evolution of our galaxy models, analyse gas and CR energy density distributions and the star formation history. In Section~\ref{sec:outflows}, we characterise the different outflow properties of our CR transport models and scrutinise how effectively the different CR momentum ranges drive galactic outflows. Finally, in Section~\ref{sec:spectra}, we explore the time dependence and spatial variation of our simulated CR spectra. We discuss extensions to our spectral CR approach in Section~\ref{sec:discussion} and conclude in Section~\ref{sec:conclusions}. Details about the variable adiabatic index and numerical test are described in Sections~\ref{sec:adiabatic-index-detail} and \ref{sec:app-tests}, respectively.

\section{Numerical Methods}
\label{sec:numerics}
\subsection{Spectral CR description and coupling to MHD}
The details of the spectral description of CRs are outlined in detail in \citetalias{GirichidisEtAl2020} and are conceptually similar to \citet{Miniati2001, YangRuszkowski2017, OgrodnikHanaszWoltanski2021}. We only highlight the main numerical features and physical processes here. We solve the Fokker-Planck equation for CRs,
\begin{align}
  \frac{\partial f}{\partial t} = & \underbrace{-\bvel\bcdot\vnabla f}_{\text{advection}} + \underbrace{\vnabla\bcdot\rkl{\tensor{D}_{xx}\bcdot \vnabla f}}_{\text{diffusion}} + \underbrace{\frac 1 3 \rkl{\vnabla\bcdot\bvel}p\frac{\partial f}{\partial p}}_{\text{adiabatic process}}\nonumber\\
  & + \underbrace{\frac{1}{p^2}\frac{\partial}{\partial p}\ekl{p^2\rkl{b_l f + D_{pp}\frac{\partial f}{\partial p}}}}_{\text{other losses and Fermi II acceleration}} + \underbrace{j}_{\text{sources}}\label{eq:FP},
\end{align}
where $f=f(\vektor{x},p,t)=\dd^6 N/(\dd x^3\,\dd p^3)$ is the isotropic part of the CR distribution function that is defined in phase space spanned by the spatial coordinates $\vektor{x}$, momentum $p=|\vektor{p}|$ and time $t$, $\bvel$ is the mean velocity of the thermal gas, $\tensor{D}_{xx}=\tensor{D}_{xx}(\vektor{x},p,t)$ is the spatial diffusion tensor, $D_{pp}=D_{pp}(\vektor{x},p,t)$ is the diffusion coefficient in momentum space, CR losses and sources are denoted by $b_l=b_l(\vektor{x},\vektor{p},t)=\dd p/\dd t$ and $j=j(\vektor{x},\vektor{p},t)$, respectively. The losses $b_l$ include the contributions from Coulomb and hadronic interactions. Here, we neglect diffusion in momentum space ($D_{pp}=0$) and therefore omit the subscript $xx$ in the spatial diffusion unless explicitly needed.

We discretise the momentum space using $N_\mathrm{spec}$ logarithmic bins, where we distinguish between the momenta at the bin faces $p_{i-1/2}$ and the bin centred values $p_i = \sqrt{p_{i-1/2}p_{i+1/2}\,}$. The particle distribution function $f$ is described by piecewise powerlaws 
\begin{align}
  f(p) &= \sum_{i=0}^{N_\mathrm{spec}}f_i(p)\\
  &= \sum_{i=0}^{N_\mathrm{spec}} f_{i-1/2} \rkl{\frac{p}{p_{i-1/2}}}^{-\slope_i}
  \theta\rkl{p-p_{i-1/2}}\theta\rkl{p_{i+1/2}-p}
\end{align}
where $f_{i-1/2}$ is the amplitude at the left face of the momentum bin, $\slope_i$ is the slope in bin $i$ and $\theta$ is the Heaviside function. The two degrees of freedom per momentum bin require two moments for the solution of the time evolution. We use the number and energy density given for each bin, which is determined by
\begin{align}
\label{eq:number-density}
  n_i &= \int_{p_{i-1/2}}^{p_{i+1/2}} 4\pi p^2 f(p) \,\dd p,\\
\label{eq:energy-density}
  e_i &= \int_{p_{i-1/2}}^{p_{i+1/2}} 4\pi p^2 f(p) T(p)\,\dd p.
\end{align}
where $T(p)=\sqrt{p^2c^2 + m_\mathrm{p}^2c^4} - m_\mathrm{p} c^2$ is the kinetic energy of individual protons and $c$ denotes the speed of light. 

We couple the Fokker-Planck equation to the MHD equations via the fluid velocity $\bvel$ of the thermal gas and the total CR pressure \citep[see equations (1) and (2) of][]{PfrommerEtAl2017} but see also \citet{HanaszStrongGirichidis2021} for a review on numerical CR models. We describe the CR population hydrodynamically by an isotropic pressure component, which is justified as long as the CRs are coupled to the thermal gas via frequent scatterings at resonant Alfv\'en waves. The CR pressure is given by
\begin{align}
\label{eq:Pcr}
P_\cra &= \int_{0}^{\infty} \frac{4\pi}{3}\, c\,p^3 \beta(p) f(p) \dd p,
%\notag\\
%  &= \int_{0}^{\infty} \frac{4\pi}{3}\, \frac{ f(p)\,p^4c^2}{\sqrt{m^2c^4+p^2c^2}}\dd p
\end{align}
where $\beta = \varv_\p/c = p/\sqrt{m_\mathrm{p}^2c^2+p^2}$ is the dimensionless velocity of the CR particle. For our spectral discretisation, the CR pressure reads
\begin{align}
P_\cra &= \sum_{i=1}^{N_\mathrm{spec}} P_\cri= \frac{4\pi}{3}\,\sum_{i=1}^{N_\mathrm{spec}}\int_{p_{i-1/2}}^{p_{i+1/2}}\,\frac{f_i(p)\,p^4c^2}{\sqrt{m_\p^2c^4+p^2c^2}}\dd p.
\end{align}

\subsection{Spatial diffusion}

\begin{figure*}
\begin{minipage}{\textwidth}
\includegraphics[width=\textwidth]{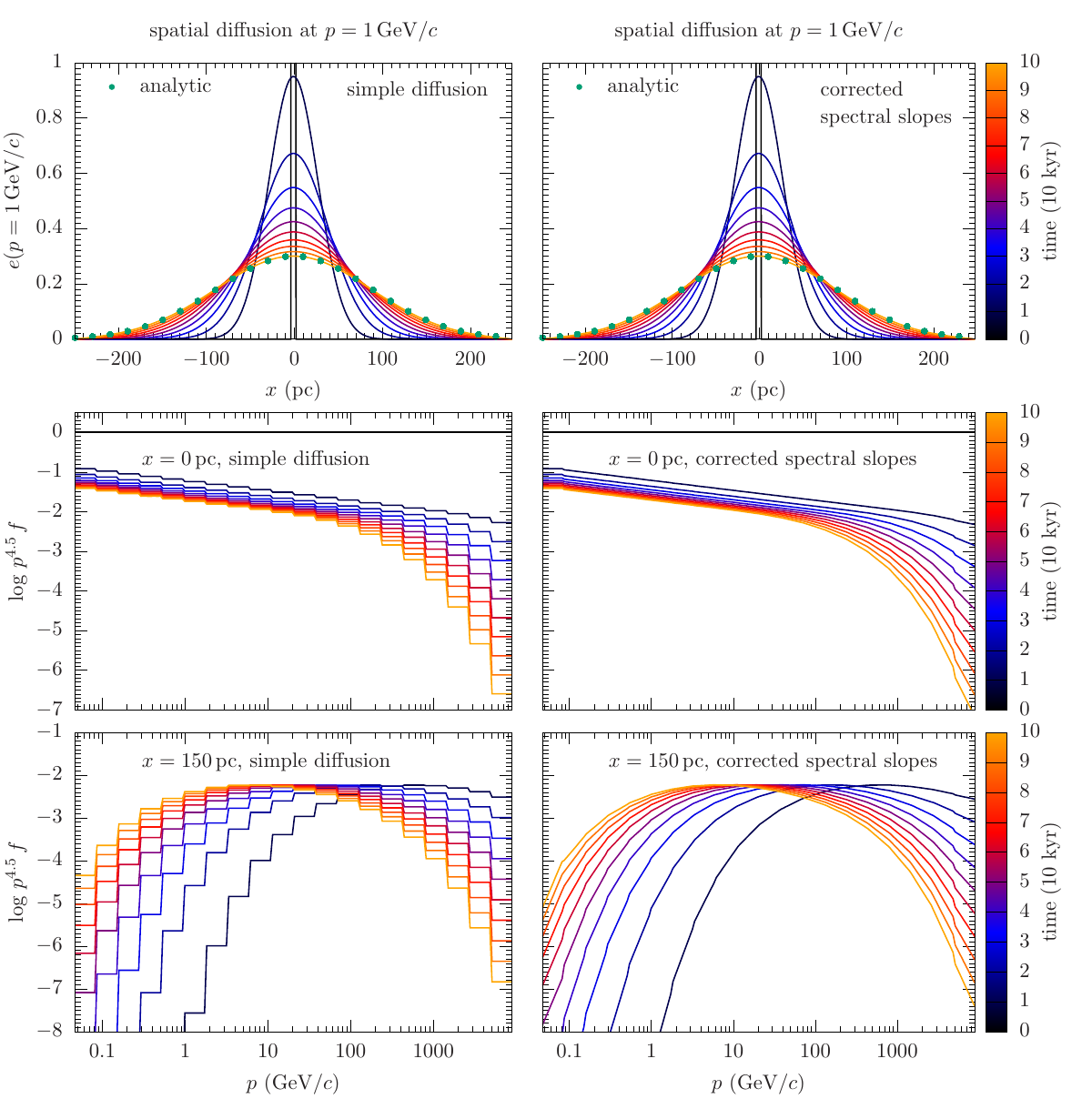}
\caption{Illustration of the effects of spatial diffusion on the spectrum. The left-hand panels correspond to spatial diffusion without spectral correction, where the initial local spectral slopes are preserved, see text. The right-hand panels show the new spatial diffusion including a spectral correction. The top plots show the spatial distribution of the CR energy density, which is identical for both methods. The middle and bottom panels depict the spectra at different positions. Using the corrected spectral slopes results in smooth spectra, while exactly conserving the energy.}
\label{fig:spectral-corr-1Ddiff-x0}
\end{minipage}
\end{figure*}

The spatial diffusion of the number density and the energy density includes the spatial derivatives of the particle distribution function. Across a momentum bin $[p_1,p_2]$, the particle diffusion operator is defined via
\begin{align}
  \label{eq:ndiff}
  \partial_t n_\mathrm{diff} &= \int_{p_1}^{p_2} 4\pi\vnabla\bcdot(\tensor{D}\bcdot\vnabla f)p^2\,\dd p\notag\\
  &=4\pi\,\vnabla \bcdot\ekl{\int_{p_1}^{p_2} p^2 \tensor{D}\bcdot\vnabla f\,\dd p}\\
  &=4\pi\,\vnabla\bcdot\rkl{\skl{\tensor{D}_n}\bcdot\vnabla n_{cr}}
\end{align}
with the individual tensor components
\begin{equation}
  \skl{D_{n,jk}} = \frac{\int_{p_1}^{p_2} p^2\,D_{jk}\,\partial_k f\,\dd p}{\int_{p_1}^{p_2} p^2 \partial_k f\,\dd p}.
\end{equation}
Similarly, for the diffusion of the energy we have
\begin{align}
  \label{eq:ediff}
  \partial_t e_\mathrm{diff} &= \int_{p_1}^{p_2} 4\pi\vnabla\bcdot(\tensor{D}\bcdot\vnabla f)p^2T(p)\,\dd p\notag\\
  &=4\pi\vnabla\bcdot \ekl{\int_{p_1}^{p_2} p^2T(p) \tensor{D}\bcdot\vnabla f\,\dd p}\\
  &=4\pi\,\vnabla\bcdot\rkl{\skl{\tensor{D}_e}\bcdot\vnabla e_{cr}},
\end{align}
with the modified diffusion coefficients
\begin{equation}
\skl{D_{e,jk}} = \frac{\int_{p_1}^{p_2} p^2T(p)\,D_{jk}\,\partial_k f\,\dd p}{\int_{p_1}^{p_2} p^2T(p) \partial_k f\,\dd p}.
\end{equation}
Ideally, one needs to compute the diffusion of number and energy density, and then reconstruct the resulting particle distribution function from the new energy and number density in each cell (in momentum and configuration space). However, this procedure is numerically difficult for several reasons. First, the spatial derivatives of $f$ include the contribution of the amplitude as well as the terms with derivatives of the slope, which makes the computation of the modified coefficients both expensive and complex. The diffusion on the spatial grid itself employs flux limiters in order to ensure that numerical differentiation in combination with directional orientation of the diffusion along magnetic field lines does not result in effective diffusion \emph{up} the gradient of the field variable. For conserved fields like the CR number and energy, limiters are conceptually intuitive. But a similar limiter needs to be applied to the distribution function and there is no analogous conserved quantity that can be limited. After the numerical diffusion step, the particle distribution function needs to be reconstructed based on the new values of $n$ and $e$. This reconstruction step is relatively sensitive to the ratio of $e/n$, which is nicely illustrated in Fig.~7 of \citetalias{GirichidisEtAl2020}. In particular for high CR momenta an inconsistent diffusion of $n$ and $e$ due to limiters and differentiation can strongly affect the reconstructed slopes. Unfortunately, those high momenta are also the spectral parts with the highest diffusion coefficients. For an implicit diffusion algorithm, the applied time step is typically set to the hydrodynamical time step, which is likely to be larger than the maximum allowed explicit diffusion time step. While implicit diffusion will be numerically stable, the solution is not guaranteed to be precise in every cell.

For all practical purposes, in which strong dynamical effects occur in neighbouring cells due to local energy injection, strong cooling (hydrodynamically as well as for CRs) and strong MHD shocks, we need a more stable workaround to the exact diffusion solution. We apply the following algorithm to overcome the numerical difficulties. We spatially diffuse the energy density $e_i$ in each spectral bin separately with diffusion coefficients based on the bin centred momenta $p_i$, $\skl{D(p)}\equiv D(p_i)$. After the diffusion step we obtain a new distribution of CR energy densities, which we use to reconstruct a particle distribution function with individual slopes, resulting in a smooth particle distribution function while conserving the energy in each spectral bin. Guaranteeing the latter, while changing the slope $\slope_i$ and the amplitude $f_{i-1/2}$ requires the number density $n$ to vary. This is justified as long as the changes are small and because we only use the number density as an auxiliary quantity in order to more accurately compute the cooling and the adiabatic process. Hence, we do not rely on a perfectly conserved number density as long as we ensure that the energy conservation in every individual spectral bin during the correction of the spectrum is not violated.

After the diffusion step we reconstruct a new logarithmic slope in bin $i$. The new slope should smoothly connect the spectral amplitude from the neighbouring lower momentum bin $i-1$ to the higher momentum bin $i+1$. Naturally, we would use the amplitudes at the bin centres $f_{i-1} = f(p_{i-1})$ and $f_{i+1}=f(p_{i+1})$. However, the amplitude in bin $i-1$, $f_{i-1}$, is connected to both, the number density $n_{i-1}$ and the energy density $e_{i-1}$. Since we have updated only the energy $e_{i-1}$ during the diffusion step we cannot reconstruct a new reliable slope directly from $n_{i-1}$ and $e_{i-1}$. Instead we use the scaling
\begin{equation}
\frac{\partial e}{\partial p} \propto fp^2 T(p) \propto fp^{2+s},
\end{equation}
where
\begin{align}
s=\frac{\partial\ln T}{\partial\ln p} = \frac{p^2}{p^2 + m_\p^2 c^2 - m_\p\sqrt{p^2c^2 + m_\p^2c^4}}.
\end{align}
For the integrated energy in each bin this yields a scaling with $e\propto fp^{3+s}$.
The logarithmic slope is defined as
\begin{equation}
q = \frac{\partial \ln f}{\partial\ln p},
\end{equation}
and the discretised slope for bin $i$ can be defined using centred derivatives, i.e.,
\begin{equation}
q_i = \frac{\ln(f_{i+1}/f_{i-1})}{\ln (p_{i+1}/p_{i-1})} = \frac{\ln(e_{i+1}/e_{i-1})}{\ln (p_{i+1}/p_{i-1})}-3-s.
\end{equation} 
For the two boundary bins we compute the one-sided derivative, namely
\begin{equation}
q_0 = \frac{\ln(e_1/e_0)}{\ln (p_1/p_0)}-3-s.
\end{equation}
for bin $0$ and similarly for the highest momentum bin.

We test the modified diffusion and correction method in a one-dimensional spatial setup with 250 cells spanning $500\,\mathrm{pc}$. The spectral grid covers momenta from 0.05 to $10^4\,\mathrm{GeV}~c^{-1}$ using 20 bins. The initial conditions are a $\delta$ distribution of CR energy in the central spatial cell at $x=0\,\mathrm{pc}$ with a spectral form of $f\propto p^{-4.5}$. We solve the diffusion equation for every spectral bin with a momentum dependent diffusion coefficient of $D(p) = 10^{28}\,\mathrm{cm^2\,s^{-1}} [p/(\mathrm{1\,GeV}~c^{-1})]^{0.5}$. The effect of the new correction method is illustrated in Fig.~\ref{fig:spectral-corr-1Ddiff-x0}. The top panels show the spatial distribution of the CR energy density over time. The coloured lines indicate the numerical solution, the dotted line shows the analytical distribution at the end.

The middle panels of Fig.~\ref{fig:spectral-corr-1Ddiff-x0} show the spectra at the centre of the box, in which we place the initial energy while the bottom panels depict the spectrum at position $x=150\,\mathrm{pc}$. In the left-hand panels we show the spectrally uncorrected diffusion, the right-hand panels use the new correction method, and the simulation time is colour-coded. In both methods the CR energy diffuses with the same accuracy by construction, so that energy conservation is not affected by the correction of the slopes. At the central position, the energy only declines with the high momentum bins diffusing faster. As a result, the spectrum steepens. Without any correction the initial slope remains unchanged throughout the evolution. After correcting the slope the spectrum retains a smooth form. Since we do not alter the energy in the corrected spectra, we conserve energy up to machine precision. In the bottom panels at $x=150\,\mathrm{pc}$ the CR energy is initially zero. The spectrum builds up over time with the highest energies first before the low energy part of the spectrum is populated. By the time the lowest momentum bin has diffused to this distance, the energy in the highest bins has already declined due to continuous diffusion. The corrected spectra are smooth in all cases.

\subsection{Adiabatic index}
\label{sec:adiabatic-index}

\begin{figure*}
\begin{minipage}{\textwidth}
\includegraphics[width=0.48\textwidth]{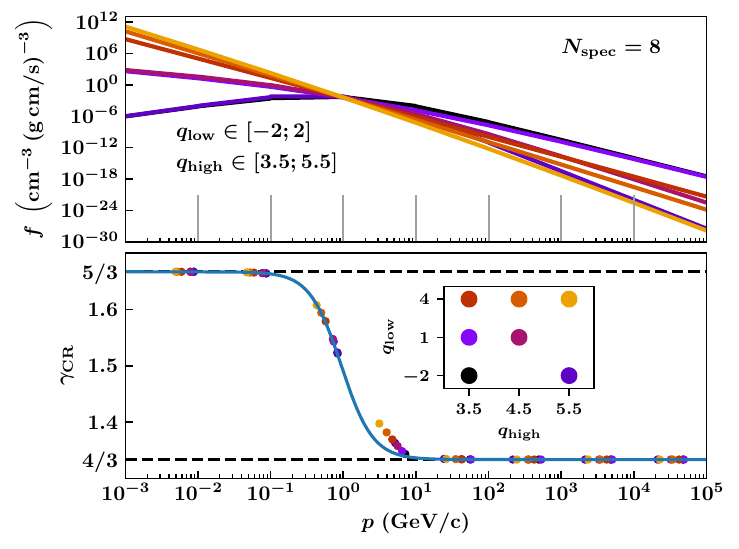}
\includegraphics[width=0.48\textwidth]{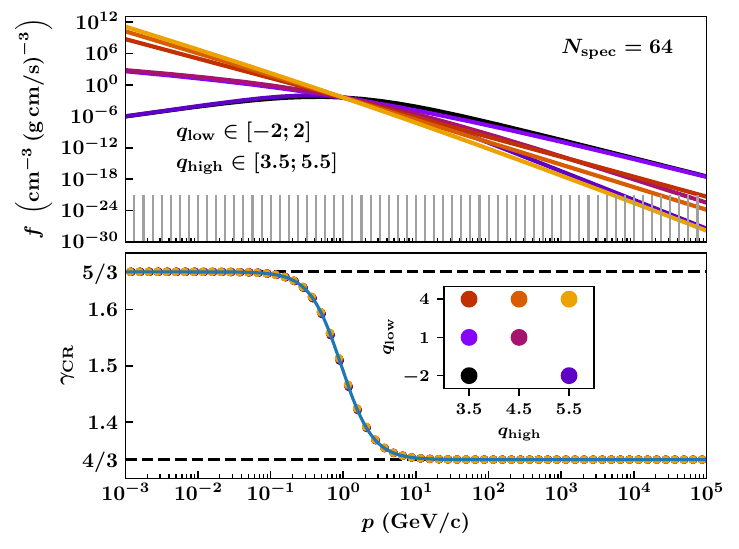}
\end{minipage}
\caption{Adiabatic index of the CRs $\gamma_\mathrm{cr}$ for different particle distribution functions and two different spectral discretizations with $N_\mathrm{spec}=8$ (left-hand side) and $N_\mathrm{spec}=64$ bins (right-hand side). For a small number of bins, the values of $\gamma_\cri$ vary with the shape of the spectra, and so do the pressure weighted average momenta of the bins. Using 64 spectral bins reduces the variations to a negligible level.}
\label{fig:gamma-CR}
\end{figure*}

In general, the adiabatic index is defined as 
\begin{equation}
\gamma \equiv \left.\frac{\mathrm{d}\ln P}{\mathrm{d}\ln\rho}\right|_S
\end{equation}
with the pressure $P$ and the density $\rho$ at constant entropy $S$. In case of a multi-component fluid, the pressure is the sum over the partial pressures. We define an adiabatic index for each component separately so that we obtain for a combined thermal and CR fluid
\begin{align}
P_\mathrm{tot} &= P_\mathrm{th} + P_\cra,\\
\gamma_\mathrm{th} &= \left.\frac{\mathrm{d}\ln P_\mathrm{th}}{\mathrm{d}\ln\rho}\right|_S,\\
\gamma_\cra &= \left.\frac{\mathrm{d}\ln P_\cra}{\mathrm{d}\ln\rho}\right|_S.
\end{align}
Hence, the effective adiabatic index is found to be the pressure weighted sum of the individual indices,
\begin{equation}
  \gamma_\mathrm{eff} = \left.\frac{\mathrm{d}\ln P_\rmn{tot}}{\mathrm{d}\ln\rho}\right|_S
  = \frac{1}{P_\mathrm{tot}}\left(\gamma_\mathrm{th} P_\mathrm{th} + \gamma_\cra P_\cra\right).
\end{equation}
Previous approaches that model the CR population only with an energy density \citep[and momentum density in the two-moment formulation,][]{ThomasPfrommer2019}, adopt a value for the adiabatic index of CRs of $\gamma_\cra=4/3$, which corresponds to an ultra-relativistic fluid \citep[e.g.][]{HanaszLesch2003, YangEtAl2012, GirichidisEtAl2016a, PfrommerEtAl2017, ButskyQuinn2018, BustardEtAl2020, ButskyEtAl2020}. If we also include low-energy CRs and resolve the CRs spectrally, the ultra-relativistic approximation is no longer valid. Since the adiabatic index alters the compressibility of the fluid, it can be relevant for the dynamics. This is particularly the case at the transition from the non-relativistic to the relativistic regime, which contains most of the CR energy. Hence, we need to compute the effective adiabatic index self-consistently. To do so, we need to account for the individual partial CR pressures and the corresponding adiabatic indices for every spectral bin. We therefore define a CR adiabatic index for each spectral bin, $i$,
\begin{equation}
\gamma_\cri \equiv \left.\frac{\mathrm{d}\ln P_\cri}{\mathrm{d}\ln\rho}\right|_S.
\end{equation}
The effective adiabatic index then reads
\begin{equation}
\gamma_\mathrm{eff} = \frac{1}{P_\mathrm{tot}}\left(\gamma_\mathrm{th} P_\mathrm{th} + \sum_{i=1}^{N_\mathrm{spec}}\gamma_\cri P_\cri\right)
\end{equation}  
where the total pressure is given by
\begin{equation}
P_\mathrm{tot} = P_\mathrm{th} + \sum_{i=1}^{N_\mathrm{spec}}P_\cri.
\end{equation}

We compute the adiabatic index as a function of the local momentum as well as for our choice of the discretized momentum bins in Appendix~\ref{sec:adiabatic-index-detail}. The adiabatic index at a localized momentum takes the simple form
\begin{align}
\label{eq:gamma-analytic-simple}
\gamma(p_0) = \frac{5}{3} - \frac{p_0^2}{3(p_0^2+m_\p^2c^2)}.
\end{align}
The solution for the discretized bins formally depends on the local choice of the binning and the shape of the distribution function. We show the adiabatic index in Fig.~\ref{fig:gamma-CR} for a variety of spectral shapes of the form
\begin{equation}
f(t=0) = A_0\,\left[\left(\frac{p}{\mathrm{1GeV/}c}\right)^{\textstyle{\frac{-q_\mathrm{low}}{w}}} + \left(\frac{p}{\mathrm{1GeV/}c}\right)^{\textstyle{\frac{-q_\mathrm{high}}{w}}}\right]^{-w}
\end{equation}
with the low-momentum slope $q_\mathrm{low}$, the high-momentum counterpart $q_\mathrm{high}$, the amplitude $A_0$, which we set to unity, and the parameter $w$ that determines the width of the transition between high and low energy part, which we also set to unity. We combine three different low-$p$ values $q_\mathrm{low}\in\{-2,\,0,\,2\}$ with three high-$p$ values $q_\mathrm{high}=\in\{3.5,\,4.5,\,5.5\}$ to illustrate the change in the adiabatic index across the spectrum, see Fig.~\ref{fig:gamma-CR}. The left- and right-hand panels depict the quantities for $N_\mathrm{spec}=8$ and $64$, respectively. The upper panels show the different spectra spanning the full momentum range from the non-relativistic to the highly relativistic limit with the spectral bin boundaries indicated by the grey vertical lines. In the lower panel we plot the adiabatic index. Besides the dots that indicate the discretized values we overplot the analytical solution of Equation~\eqref{eq:gamma-analytic-simple}. For the discretized formulation the adiabatic index and the average momentum of each momentum bin depend on the shape of the particle distribution function, i.e. on the local slope in bin $i$. We therefore plot the adiabatic index in each bin as a function of the pressure-weighted average momentum in each bin, which explains the differently placed points along the abscissa. The fact that the adiabatic index at infinite spectral resolution does not depend any more on the details of the spectral shape is well reflected in the high-resolution example with 64 bins. The variations in the spectra are not visible in $\gamma_\mathrm{CR}$, which is very close to the analytic formula.

\section{Galaxy simulation setup}
\label{sec:code}

\begin{table}
\caption{Spectral bins and the corresponding diffusion coefficients. Listed are the bin number, the left-hand momentum boundary, the central momentum as well as the corresponding diffusion coefficient.}
\label{tab:spec-bins}
\begin{tabular}{cccc}
\hline
$i$ & $p_{i-1/2}$ & $p_i$ & $D$\\
  & (GeV/$c$) & (GeV/$c$) & ($10^{28}$\,cm$^{2}$\,s$^{-1}$)\\
  \hline
0   &   0.1   &  0.178   &   0.422\\
1   &   0.316&  0.562  &   0.750\\
2   &   1.0   &  1.78     &   1.33\\
3   &   3.16 & 5.62      &   2.37\\
4   &   10.0      & 17.8     &   4.22\\
5   &   31.6   & 56.2        &   7.50\\
6   &   100                       &  178                            &   13.3\\
7   &   316                       & 562                             &   23.7\\
\hline
\end{tabular}
%\medskip
%Listed are the bin number, the left-hand momentum boundary, the central momentum as well as the corresponding diffusion coefficient.
\end{table}

We link the spectral CR solver to the \textsc{Arepo} code, which decomposes the computational domain into an unstructured moving mesh that is second-order accurate in space and time \citep{Springel2010,Pakmor2016a,WeinbergerSpringelPakmor2020}. The CR quantities are coupled to the combined CR-MHD solver described in \citet{PfrommerEtAl2017} in the advection-diffusion approximation. The magnetic field is evolved using the ideal MHD approximation \citep{PakmorBauerSpringel2011,PakmorSpringel2013}. The anisotropic diffusion equation for each spectral bin is solved using the implicit diffusion solver described in \citet{PakmorEtAl2016}.

We simulate isolated dwarf galaxies similar to those in \citet{PakmorEtAl2016} and \citet{PfrommerEtAl2017}. The total gas mass of the simulated galaxies is $1.55\times10^{10}\,\mathrm{M}_\odot$, embedded in a dark matter halo with a total mass of $10^{11}\,\mathrm{M}_\odot$, i.e. a baryon mass fraction of $\Omega_\mathrm{b}/\Omega_\mathrm{m}=0.155$. Initially, the gas follows the NFW dark matter profile \citep{NavarroFrenkWhite1997}, for which we adopt a concentration of $c_{200} = r_{200}/r_\mathrm{s} = 7$, where $r_\mathrm{s}$ is the characteristic scale radius of the NFW profile and $r_{200}$ is the radius that encloses 200 times the critical density of the universe. The gas and the dark matter halo are in slow rotation with the same initial angular frequency. The rotation speed is given implicitly using the dimensionless spin parameter $\lambda=J|E|^{1/2}/(GM_{200}^{5/2})=0.05$ with the angular momentum $J$, the total halo energy $|E|$, Newton's constant $G$ and the mass within $r_{200}$, $M_{200}$. The halo contains in total $10^6$ gas cells with a target mass of $1.55\times10^4\,\mathrm{M}_\odot$. The adaptive mesh refinement of the Voronoi cells ensures that adjacent cells differ by at most a factor of ten in volume.

\begin{figure*}
\begin{minipage}{\textwidth}
\includegraphics[width=\textwidth]{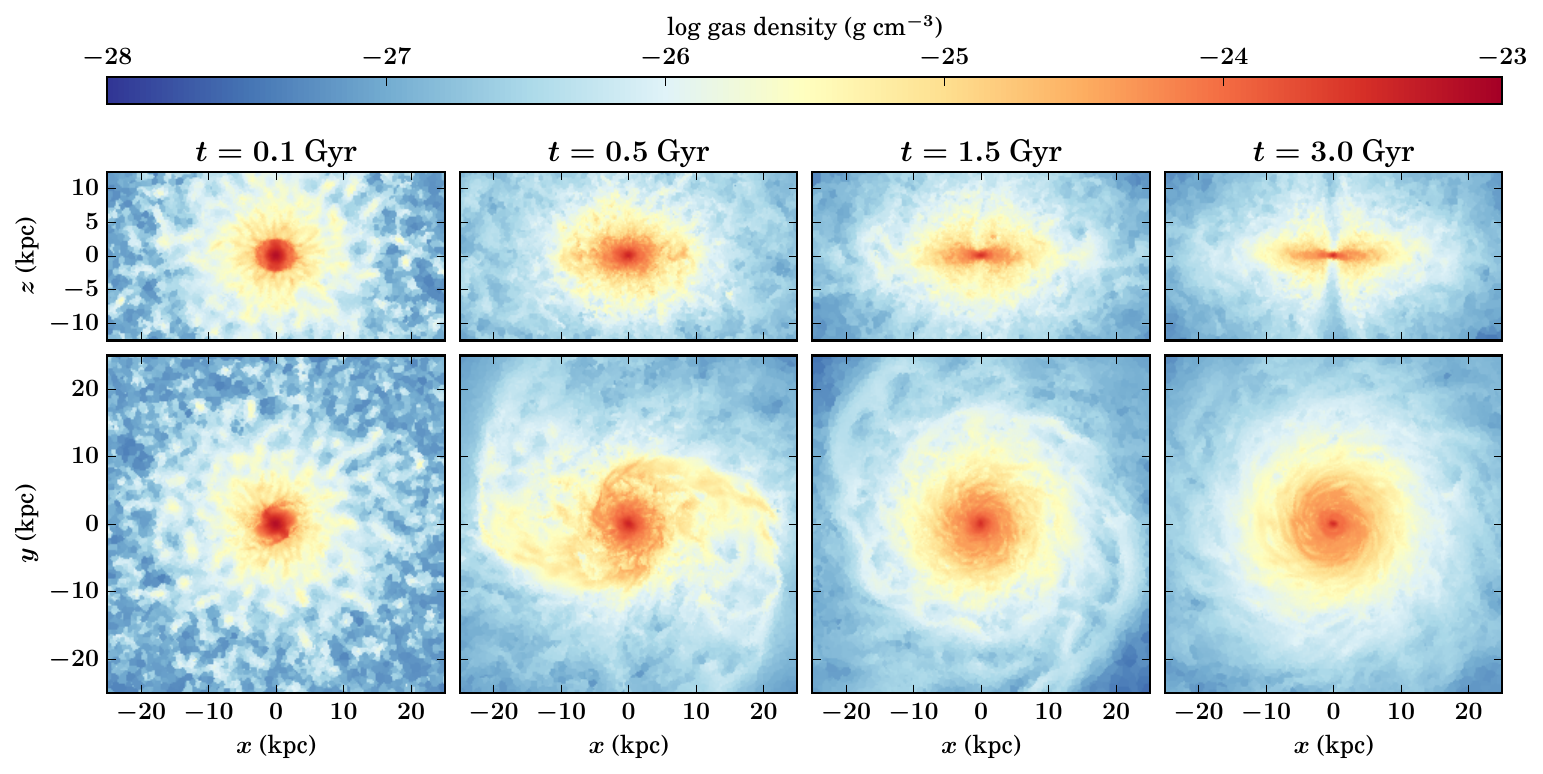}
\caption{Time evolution of the gas density in simulation \spec in cuts through the center of the simulation box. The upper panels show the galaxy edge-on. The bottom panels show the face-on counterparts. After $t=0.1\,\mathrm{Gyr}$ the galaxy has an almost spherical structure with a strong concentration of gas in the central region. At $t=0.5\,\mathrm{Gyr}$ the angular momentum causes the galaxy to flatten and a disc forms. At the later stages the disc structure becomes more apparent and a vertical underdense cone emerges above and below the disc as a consequence of strong vertical outflows from the centre.}
\label{fig:spectral-gas-time-evol}
\end{minipage}
\end{figure*}

The thermodynamics of the gas includes radiative cooling and star formation based on the pressurized ISM model in \citet{SpringelHernquist2003}. In this subgrid ISM model, star formation is included in a stochastic manner following the observed relation between gas surface density and star formation rate surface density \citep{KennicuttSchmidt1998}. The model assumes the hot and cold phase of the ISM to be in pressure equilibrium and parametrizes the overall effective pressure of the ISM in form of a stiff equation of state that implies a minimum pressure. The density threshold of star formation is set to $\rho_0 = 4\times10^{-25}\,\gpcc$. Numerically, star formation is modelled by injecting a star particle into the simulation with instantaneous stellar feedback. In addition to the implicite subgrid thermal feedback, CRs are injected with an efficiency of $10$ per cent of the canonical SN energy of $10^{51}\,\mathrm{erg}$ \citep{HelderEtAl2012,MorlinoCaprioli2012,AckermannEtAl2013}. We note, however, that the SN energy can differ perceptibly from the canonical value and that the local CR acceleration efficiency could also be as low as 5 per cent \citep{PaisEtAl2018}. We do not include any CRs in the initial conditions, so all CR energy is injected in connection to star formation.\footnote{Note that we do not use the wind particles of \citet{SpringelHernquist2003} in our simulations presented here.}

We perform four different simulations. Three of them use the spectrally grey CR approach as described in \citet{PakmorEtAl2016, PfrommerEtAl2017}. The physical processes include cooling based on a spectral steady state assumption, CR advection and anisotropic diffusion. Using the grey approach we run one simulation in which we only consider CR advection (\greyadv). In addition to advection, two further simulations include anisotropic diffusion with diffusion coefficients parallel to the magnetic field lines of $D=10^{28}\,\mathrm{cm^2\,s^{-1}}$ (\greydiff) and $D=4\times10^{28}\,\mathrm{cm^2\,s^{-1}}$ (\greydiffhi), respectively. The fourth simulation (\spec) uses the new spectral method with eight spectral bins ranging from $100\,\mathrm{MeV}~c^{-1}$ to $1\,\mathrm{TeV}~c^{-1}$ equally spaced in logarithmic scale. We inject CRs with the spectral form of a power law $f(p)\propto p^{-4.5}$ into the local environment of every newly created star particle using a spherical top hat filter that contains the closest 32 mesh cells of the star. We transport the CR distribution in configuration space via advection and (momentum-dependent) diffusion, and in momentum space we model adiabatic processes, Coulomb and hadronic cooling. The spatial diffusion parallel to the magnetic field lines is given by
\begin{equation}
\label{eq:diff-p-scaling}
D(p) = 10^{28}\,\rkl{\frac{p}{1\,\mathrm{GeV}~c^{-1}}}^s\,\mathrm{cm^2\,s^{-1}},
\end{equation}
where we adopt the scaling of the diffusion coefficient $s=0.5$ as inferred from observed beryllium isotope ratios \citep{Evoli2020}. As described in the diffusion method above we use the diffusion coefficients at bin centres, which are listed in Table~\ref{tab:spec-bins}. We run all simulations for a total time of $3\,\mathrm{Gyr}$.

\section{Global evolution}
\label{sec:global}

\begin{figure*}
\begin{minipage}{\textwidth}
\begin{center}
\includegraphics[width=0.8\textwidth]{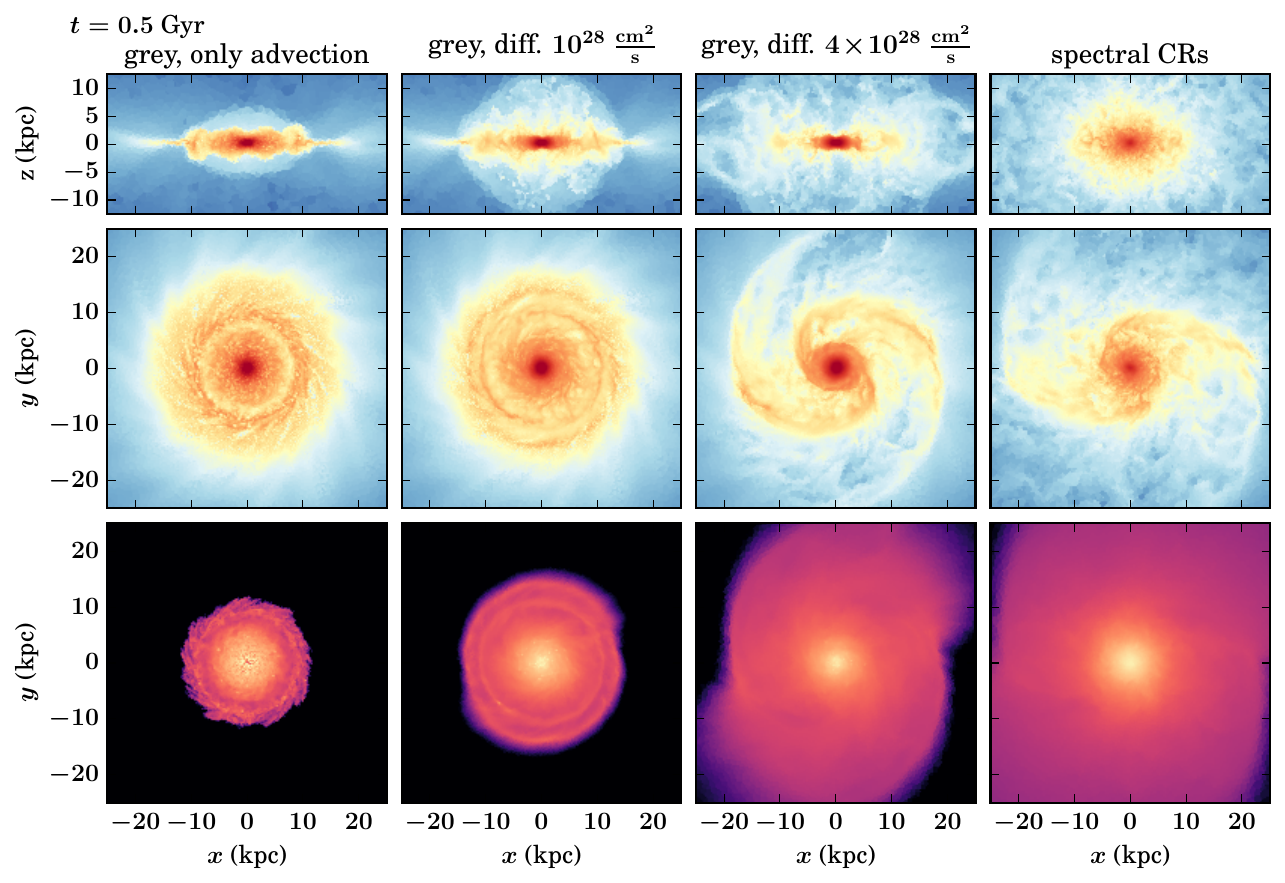}\\
\includegraphics[width=0.8\textwidth]{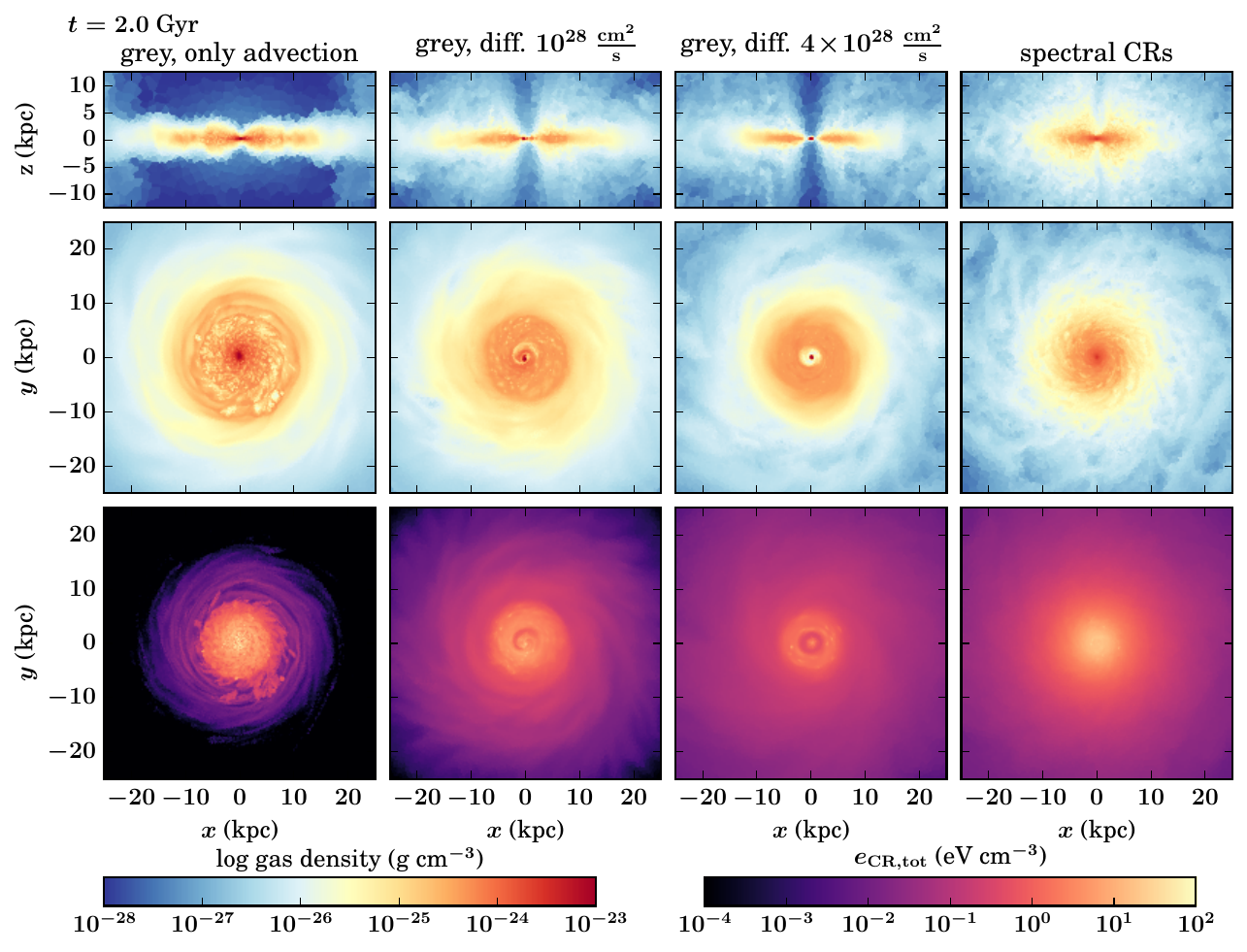}
\end{center}
\caption{Differences in the gas morphology for all runs at two times, $t=0.5\,\mathrm{Gyr}$ (top) and $t=2\,\mathrm{Gyr}$ (bottom). From left to right we show the simulations \greyadv, \greydiff, \greydiffhi, and \spec in which the diffusion coefficient scales as $D\propto p^{0.5}$, see Equation~\eqref{eq:diff-p-scaling}. In the advection run the CRs remain relatively close to the injection region. Diffusion of the CR energy density with a constant diffusion coefficient (two middle panels) drive outflows early on (top panels at $t=0.5\,\mathrm{Gyr}$). The higher diffusion coefficient can push gas further out and forms a strong spiral structure. The spectral run (right-hand panels) drives even stronger outflows while keeping the central region of the galaxy more spherical. At late times the grey CR models clear out a vertical cone from where the outflow is launched. This effect is less pronounced in the spectral CR run.}
\label{fig:density-comparison-all-models}
\end{minipage}
\end{figure*}

\begin{figure*}
\begin{minipage}{\textwidth}
\includegraphics[width=0.45\textwidth]{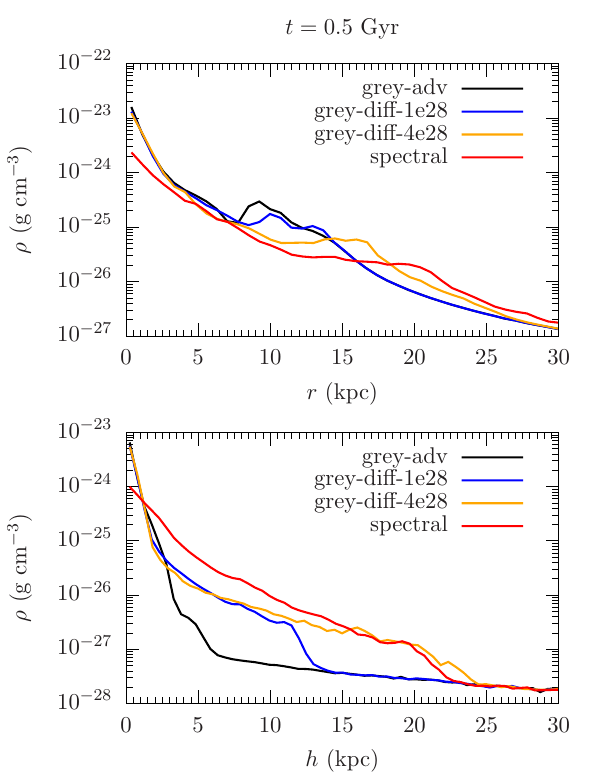}
\includegraphics[width=0.45\textwidth]{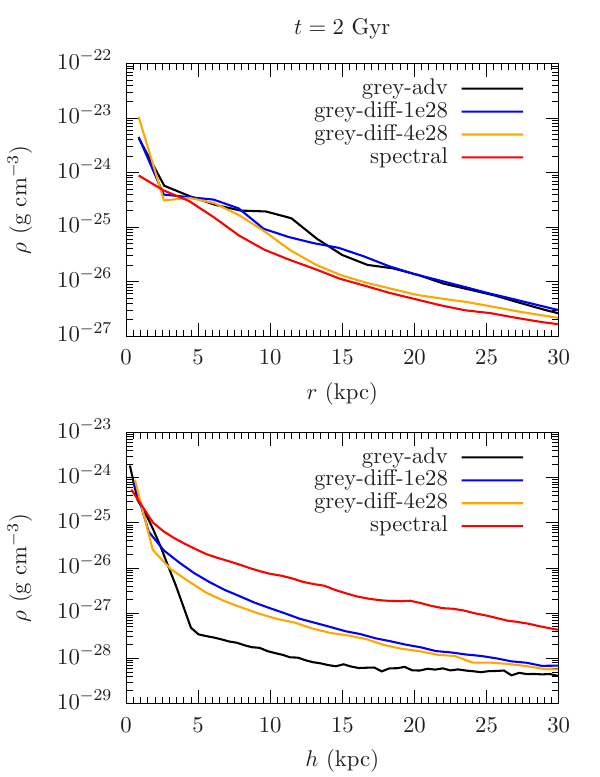}
\caption{Radial profiles in a cylinder of height $\pm1\,\mathrm{kpc}$ centred on the midplane (top) and vertical profiles in a galactic-centric cylinder of radius $5\,\mathrm{kpc}$ (bottom). The left- and right-hand panels show simulation times $t=0.5\,\mathrm{Gyr}$ and $2\,\mathrm{Gyr}$, respectively. The spectral model shows the lowest central densities. Outside the central region the radial profiles overall show minor differences between the models. In the case of the vertical profiles the spectral model starts to push the gas further out at an earlier time. The two grey diffusion models differ mainly at $t=0.5\,\mathrm{Gyr}$, where faster diffusion transports gas to larger heights. At later times there is no significant difference between the grey diffusion runs, whereas the spectral run shows significantly larger densities at all heights. If only grey advective transport is included the profiles show the lowest circum-galactic density and do not change much over time.}
\label{fig:density-profiles-all-models}
\end{minipage}
\end{figure*}

\begin{figure}
\centering
\includegraphics[width=8cm]{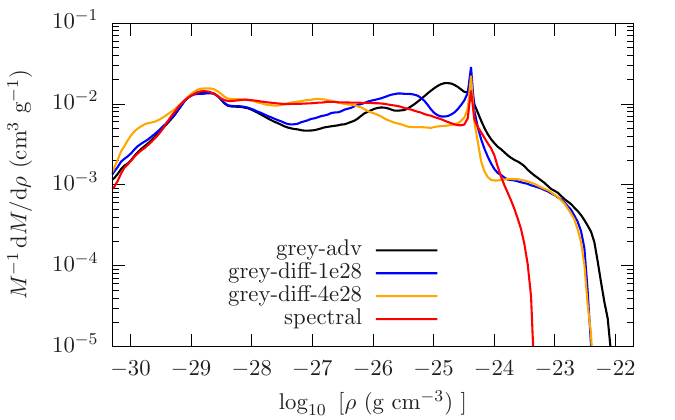}
\caption{Probability density function of the gas density for all simulations time averaged from $t=0.5-3\,\mathrm{Gyr}$. The maximum density in the spectral CR model is approximately one order of magnitude smaller than in the other models. This results in slower cooling in the disc and fewer star forming regions.}
\label{fig:density-pdf-time-avg}
\end{figure}

\begin{figure}
\includegraphics[width=0.45\textwidth]{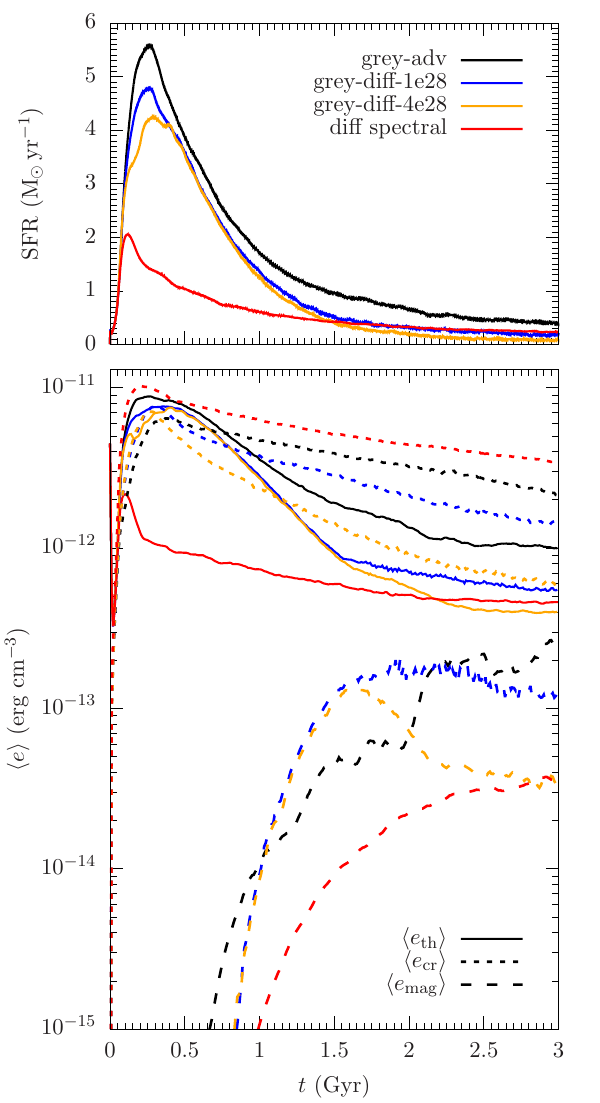}
\caption{Star formation rate (top) and volume-weighted energy densities in the disc (bottom) over time. The overall temporal evolution of the star formation rate is similar in all runs. The spectral CR model forms stars at a noticeably lower rate with a peak less than half of the rates achieved by the two grey diffusion runs. After about $1.5\,\mathrm{Gyr}$ the star formation rates of the spectral and the two grey diffusion runs are comparable. The advection run forms most stars over the entire simulation time. The energy densities (measured in a cylinder with radius $10\,\mathrm{kpc}$ and height $\pm0.5\,\mathrm{kpc}$ around the centre) directly reflect the star formation rate in the thermal gas while the CR energy density differs as a result of cooling and diffusion. The magnetic energy density is slowly increasing, and saturates more than an order of magnitude below the thermal and CR energy density.}
\label{fig:sfr-energies-time-evol}
\end{figure}

The dynamical evolution differs between the individual setups. However, there are a few common features in the evolution for all models. In all setups, the initially rigidly rotating gas cloud cools fastest in the centre, loses pressure support and starts forming stars soon after the simulation starts. The initial burst of star formation proceeds inside out and the injected CR energy peaks in the central region. Over time, the angular momentum forms a disc and the initial bursty mode of star formation changes to a simmering mode. The injected CRs partially cool and are partially transported away from the injection region by advection and/or diffusion. The resulting CR pressure gradient then launches outflows from the centre of the galaxy. Figure~\ref{fig:spectral-gas-time-evol} illustrates the time evolution of simulation \spec. Shown is the density in cuts through the centre of the domain edge-on (top panels) and face-on (bottom panels) at four different evolutionary stages. One clearly notices the formation of a spiral arm structure followed by a flattening to form a disc. At the end of the simulations all models including diffusion form a cone of low-density gas from the centre, which is created by outflows. 

The differences in the gas and CR energy distribution between the runs is depicted in Fig.~\ref{fig:density-comparison-all-models}. We show the density distribution in cuts through the centre of the box for two different times and two different orientations (top two panels) as well as the face-on CR energy density (bottom panels). The two different panels correspond to two different times. The advection run does not launch a galactic-scale outflow and instead drives fountain flows that facilitate mixing. All diffusion models drive galactic-scale outflows but the onset of those differs: we observe an earlier launching of outflows through the sequence of simulations \greydiff to \greydiffhi to \spec. The spectral model launches outflows from the entire star-forming disc more spherically in comparison to the grey diffusion models. This also results in a low-density cone from the centre, which forms later and is less pronounced in comparison to the grey diffusion models. The fast diffusion of high-energy CRs results in a more spatially extended CR distribution in the \spec model in comparison to all other models.

We quantitatively analyse the difference between the runs using gas density profiles in Fig.~\ref{fig:density-profiles-all-models}. The top panels show the radial density profiles in a flat cylinder of height $\pm1\,\mathrm{kpc}$ above and below the midplane of the disc, the bottom panels show vertical profiles around the centre of the galaxy within a cylinder of radius $5\,\mathrm{kpc}$. Overall the vertical distributions differ more strongly between the models in comparison to the radial profiles. The spectral model is more efficient in transporting gas to larger heights above the disc in comparison to the grey diffusion models, which behave very similarly after an initial phase. The advection run shows the lowest vertical density profiles over all times. We note that the central densities are significantly lower in the spectral CR model in comparison to all other models, except for the vertical profiles at late times.

We further analyse the gas density using the time averaged ($0.5-3\,\mathrm{Gyr}$) probability density function (PDF), depicted in Fig.~\ref{fig:density-pdf-time-avg} for all simulations. In the low-density regime the PDFs do not differ perceptibly. Up to a density of $\rho\sim10^{-24}\,\mathrm{g\,cm}^{-3}$ the relative difference between the runs varies by a factor of a few at most. Above $\rho\sim10^{-24}\,\mathrm{g\,cm}^{-3}$, the PDF of the spectral CR model steeply declines with a maximum density of a few times $10^{-24}\,\mathrm{g\,cm}^{-3}$. The two grey diffusion runs reach an order of magnitude higher densities, which is exceeded by the grey advection run with a maximum value of $\approx10^{-22}\,\mathrm{g\,cm}^{-3}$. A similar effect was found by \citet{SemenovKravtsovCaprioli2021}. Their model including a suppression of CR diffusion in star forming regions results in stabilizing CR pressure gradients, lower the maximum densities.

The lower maximum density directly translates into a lower star formation rate as indicated in the top panel of Fig.~\ref{fig:sfr-energies-time-evol}. All models exhibit an overall similar behaviour, in which the initial cooling phase of the gas and the fast central collapse lead to a burst of star formation after the first $100-300\,\mathrm{Myr}$. After this initial phase the star formation rate exponentially declines. The spectral CR model shows a much smaller burst, which is less than half of that of any other model at the peak. This indicates that run \spec~is much more efficient in quenching star formation, which is in line with the lower central densities found in Fig.~\ref{fig:density-profiles-all-models} and the density PDF in Fig.~\ref{fig:density-pdf-time-avg}. The lower panel of Fig.~\ref{fig:sfr-energies-time-evol} shows the energy densities in the disc, measured in a flat cylinder with a radius of $10\,\mathrm{kpc}$ around the centre and a height of $\pm0.5\,\mathrm{kpc}$ around the mid-plane. The magnetic energy is initialized as a small seed field with a strength of only $10^{-10}\,\mu\mathrm{G}$ and needs to be amplified via compression and a turbulent dynamo \citep{PfrommerEtAl2021}. While these amplification processes cause the magnetic energy density to saturate at the gravo-turbulent energy density, they are insufficient to saturate the magnetic energy at the thermal or CR energies in a galaxy of halo mass $10^{11}~\rmn{M}_\odot$. Instead, additional magnetic amplification mechanisms are required to grow the field further.

The thermal energy density directly reflects the star formation rate, i.e. the amount of thermal energy directly scales with the injected thermal SN energy. This means that the thermal energy in the disc cools efficiently and is radiated away as soon as the thermal supply by the star formation rate drops. The CR energy density also rapidly increases with the star formation rate. However, contrary to the thermal energy, the cooling of CRs is less efficient and a larger fraction of the CR energy density remains in the disc after the initial burst of star formation.

The spectral model reveals a remarkable CR-to-thermal energy density ratio. Because of the smallest star formation rate of all models, the generated thermal and CR energy is smallest. However, the spectral model ends up with the largest amount of CR energy. Responsible for this effect are the low-energy CRs that neither diffuse fast enough out of the disc region nor do they efficiently cool. While low-energy CRs with momenta smaller than GeV~$c^{-1}$ cool efficiently via Coulomb losses in the cold and dense phase of the ISM, the densities in the pressurized ISM of our simulations are relatively low so that the net cooling of the spectral CRs is smaller in comparison to the cooling of the assumed underlying steady state CR distribution used in all our grey CR approximations. We will investigate the detailed spectra in Section~\ref{sec:spectra}.

\section{Outflows}
\label{sec:outflows}

\begin{figure*}
\begin{minipage}{\textwidth}
\centering
\includegraphics[width=0.9\textwidth]{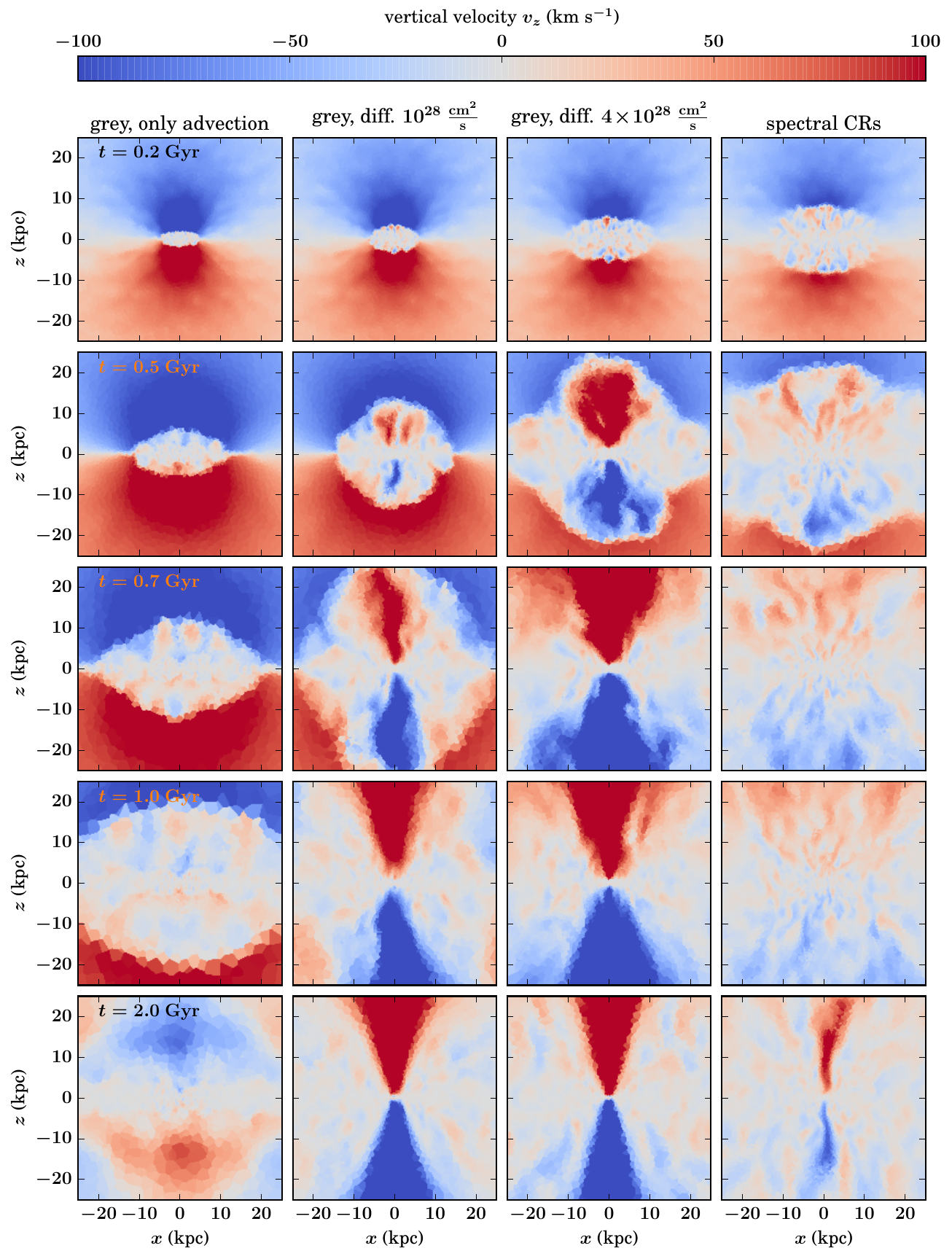}
\caption{Vertical velocity slices for different times from 0.2 (top) to 2\,Gyr (bottom) for all simulated models from left to right. The spectral model (right-hand panels) is more efficient both in driving the early outflow front as well as driving outflows from different regions of the disc. The grey diffusion runs (two middle panels) quickly form a low-density outflow cone that emerges from the centre while the spectral model forms this cone only after 2\,Gyr. The advection run (left-hand panels) only produces CR-driven fountain flows that cause mixing rather than outflows.}
\label{fig:velz-all-models}
\end{minipage}
\end{figure*}

\begin{figure}
\includegraphics[width=0.45\textwidth]{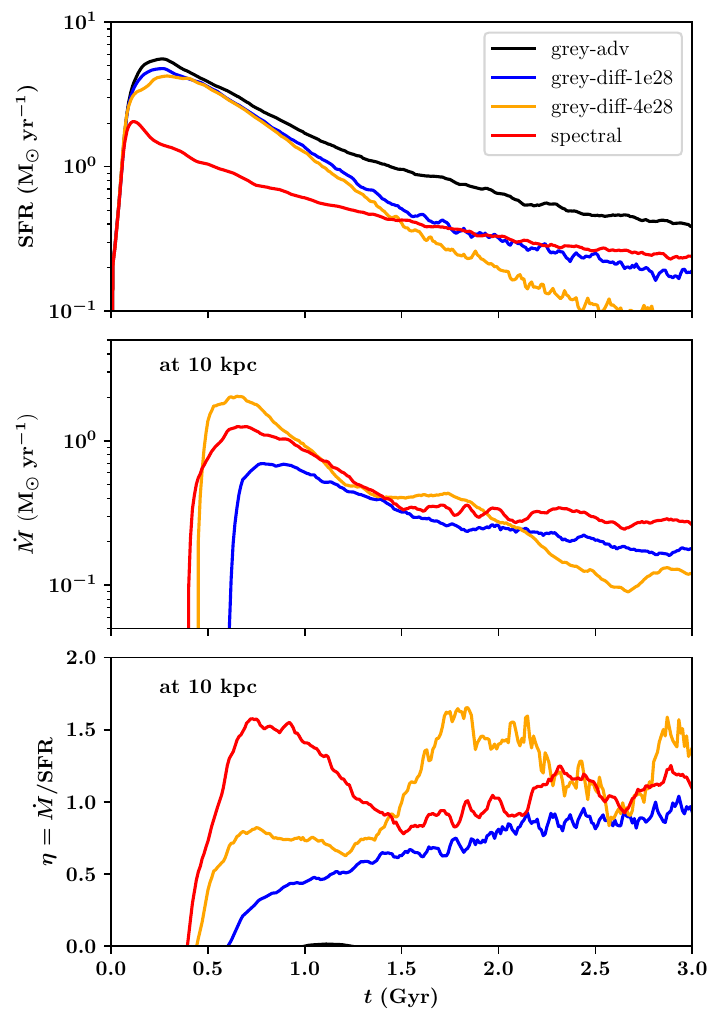}
\caption{Outflow properties over time. Shown are from top to bottom the star formation rate, the outflow rate measured at a height of $10\,\mathrm{kpc}$ as well as the mass loading factor $\eta=\dot{M}/\mathrm{SFR}$. Despite the lowest star formation rate and the resulting lowest CR injection, the spectral model develops the largest mass loading factor, which indicates that the spectral model is more efficient in influencing the outflow dynamics. Over time the mass loading factors approach each other and are comparable. The grey advection run does not drive gas out to a distance of $10\,\mathrm{kpc}$.}
\label{fig:outflows-time-evol}
\end{figure}

\begin{figure*}
\begin{minipage}{\textwidth}
\includegraphics[width=\textwidth]{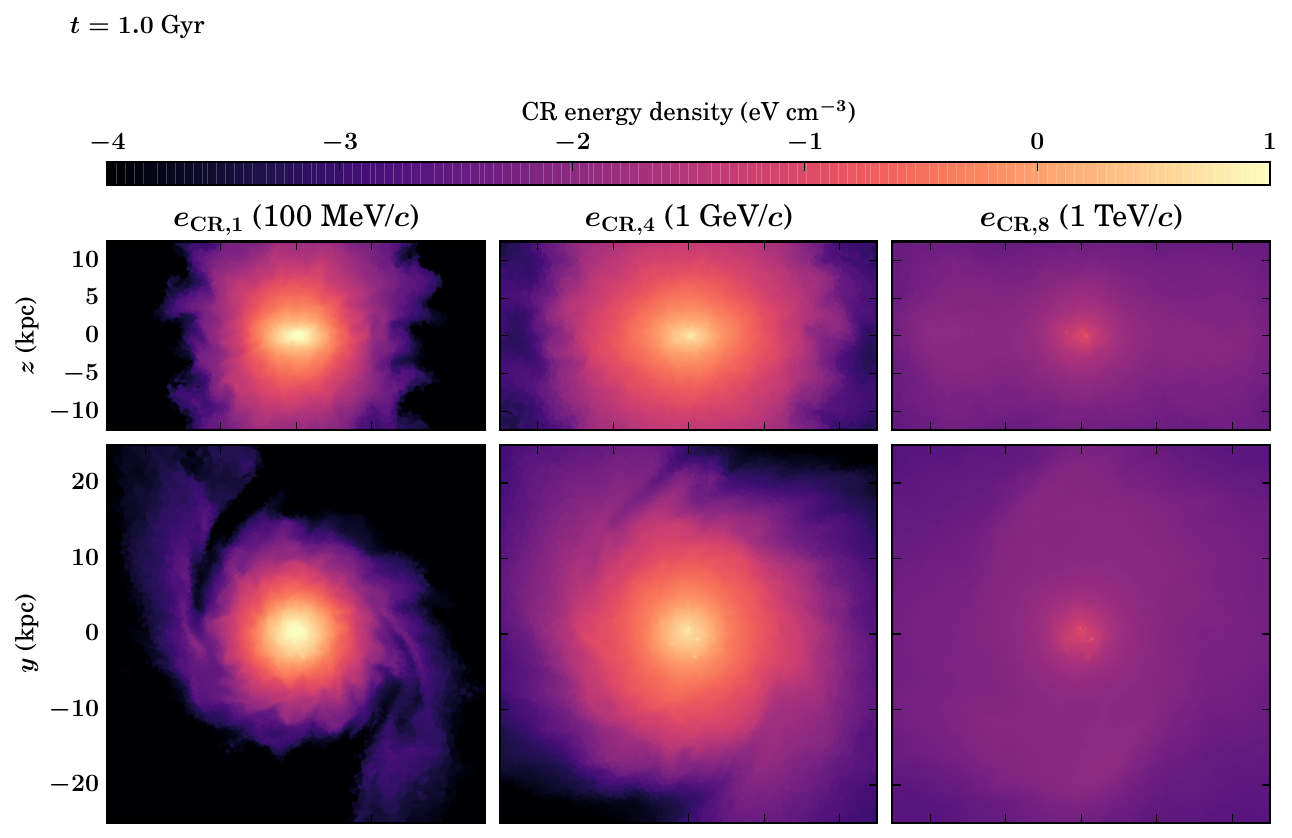}
\includegraphics[width=\textwidth]{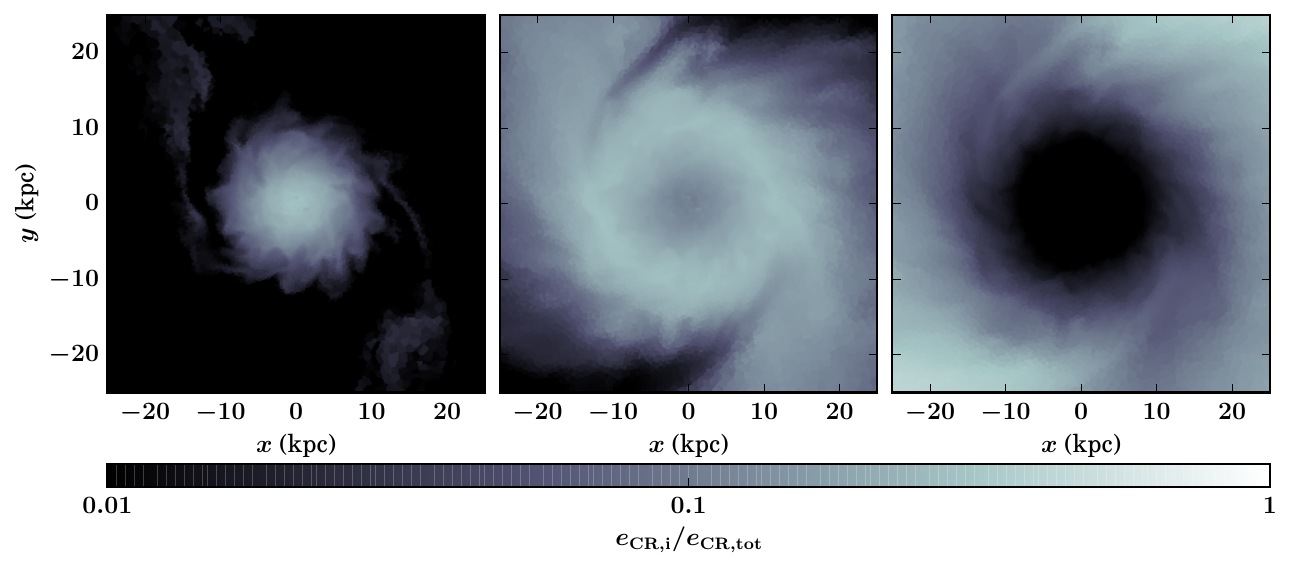}
\caption{Distribution of CR energy density after $1\,\mathrm{Gyr}$ of evolution in cuts through the centre of the galaxy edge-on (top panels) and face-on (middle panels). Shown are from left to right the lowest CR momentum bin (100\,MeV~$c^{-1}$), the intermediate bin (1 GeV~$c^{-1}$), and the highest momentum bin (1 TeV~$c^{-1}$). In addition, the bottom panels show the ratio of the energy density in each individual momentum bin to the total CR energy for face-on cuts. The higher the energy, the faster the diffusive transport and the smaller the CR energy density contrast. The ratios highlight that the centre is dominated by MeV CRs. Most of the disc is dominated by GeV CRs and in the outer regions only TeV CRs are present.}
\label{fig:diffusion-different-energies}
\end{minipage}
\end{figure*}

\begin{figure*}
\begin{minipage}{\textwidth}
\includegraphics[width=0.5\textwidth]{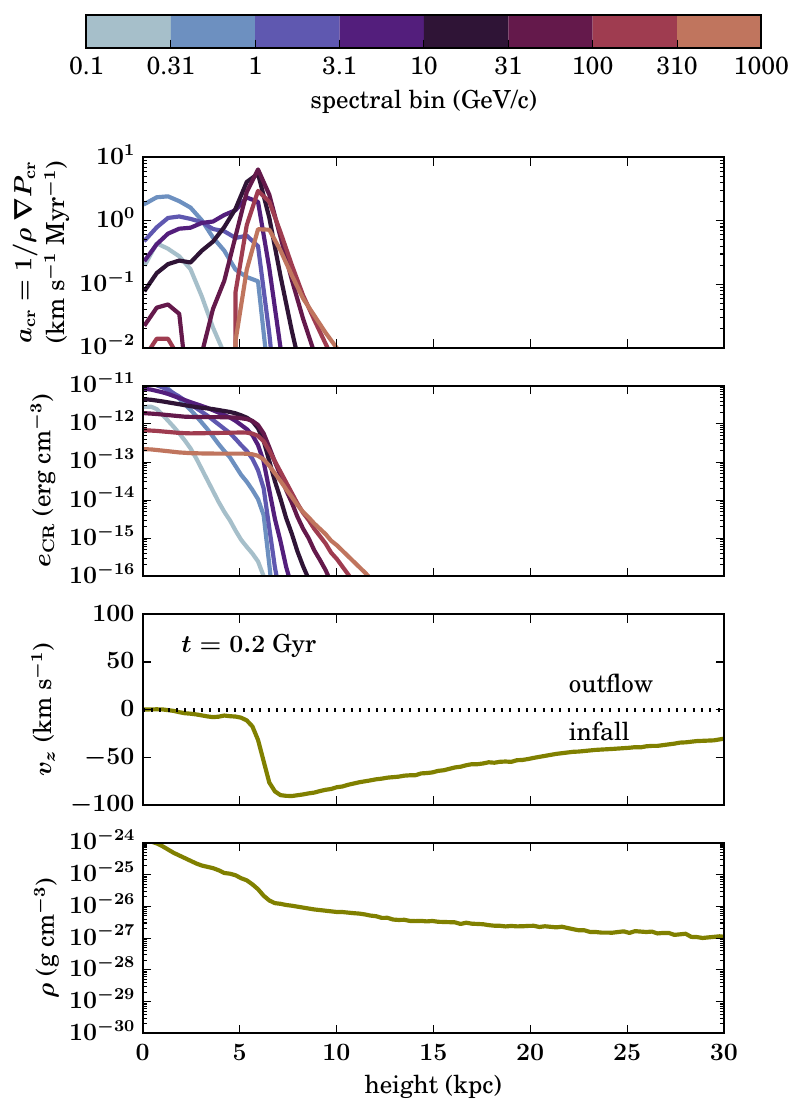}
\includegraphics[width=0.5\textwidth]{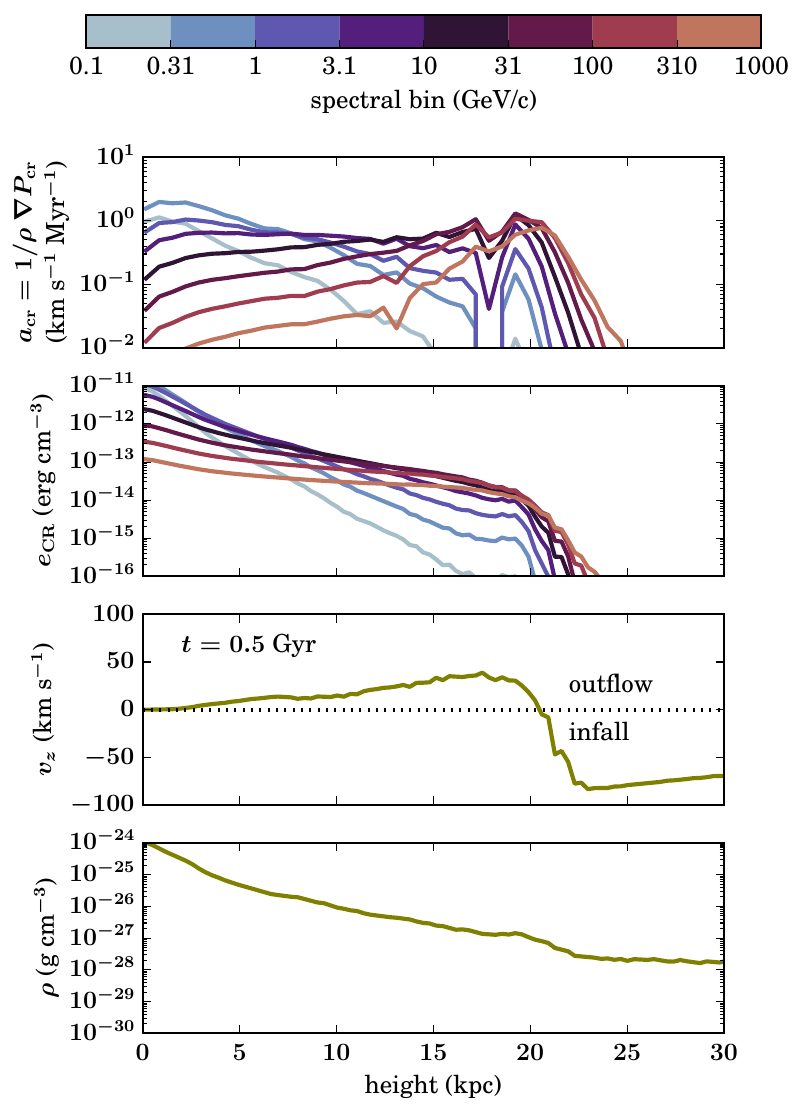}\\
\caption{Vertical profiles computed as radial averages in a cylinder of radius $r=5\,\mathrm{kpc}$ centred on the galactic centre. From bottom to top we show the gas density, the vertical velocity $\varv_z$, the spectrally resolved CR energy density, and the spectrally resolved acceleration due to the CR pressure gradient. The left-hand panels are at $t=0.2\,\mathrm{Gyr}$, the right-hand ones at $t=0.5\,\mathrm{Gyr}$. The different CR momenta are colour coded. The prominent change in velocity and density at approximately $z=6\,\mathrm{kpc}$ ($0.2\,\mathrm{Gyr}$) and $20\,\mathrm{kpc}$ ($0.5\,\mathrm{Gyr}$) mark the first CR-driven front that halts the infall of halo gas from the circum-glactic medium onto the disc. This front is driven by CRs with a momentum of $\sim30-100\,\mathrm{GeV}~c^{-1}$, which is well reflected as peaks in the acceleration profiles. At $t=0.5\,\mathrm{Gyr}$ the CRs did not only halt the infall, but also started driving an outflow (positive $\varv_z$ at $z < 20\,\mathrm{kpc}$), which is mainly supported by the pressure contribution from CRs with lower momenta ($\lesssim10\,\mathrm{GeV}~c^{-1}$), in line with the acceleration profiles, see also Fig.~\ref{fig:peak-momentum-diff-heights-time-evol} and the explanation in the text.}
\label{fig:profiles-including-spectral-energies}
\end{minipage}
\end{figure*}

\begin{figure}
\includegraphics[width=0.45\textwidth]{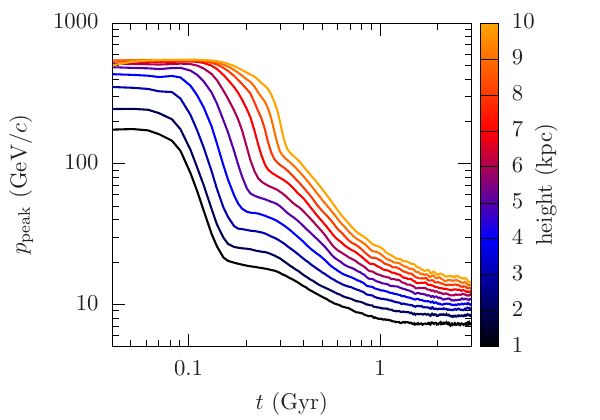}
\caption{Kinetic energy-weighted CR momentum $p_\mathrm{peak}$ as a function of simulation time for different heights above the disc (colour coded). The peak momentum reflects the dominant CR momentum in the spectrum that is able to provide most of the energy for driving outflows. The monotonic behaviour of $p_\mathrm{peak}$ with height is connected to the faster diffusion speeds for higher momenta and the resulting dominance of larger momenta in the spectrum. Over time, the first outflow front, which halts the infall of gas from the halo is accompanied by a decrease in $p_\mathrm{peak}$ between $t\sim0.1-0.4\,\mathrm{Gyr}$. Strong outflows driven at later times are supported by lower and lower CR momenta that approach $p_\mathrm{peak}\approx10\,\mathrm{GeV}~c^{-1}$ at late times.}
\label{fig:peak-momentum-diff-heights-time-evol}
\end{figure}

In this section we focus on the outflow of gas from the galaxies driven by star formation processes. Relating star formation and outflows requires a word of caution since the instantaneous star formation in the densest regions of the galactic disc does not immediately correlate with outflows at a certain height. One particularly critical measure is the mass loading factor $\eta=\dot{M}_\mathrm{out}/\mathrm{SFR}$. i.e. the amount of outflowing gas mass per unit mass that forms stars. The delay between star formation and the corresponding outflows depends on a number of factors like the thermal state in the ISM that drives the outflow \citep[e.g.,][]{WalchEtAl2015, GattoEtAl2017, Hu2019, RathjenEtAl2021}, the clustering of feedback injection \citep[e.g.][]{GirichidisEtAl2016b, FieldingQuataertMartizzi2018}, the transport speed at which the different phases travel from the star formation site to the height at which we measure the outflow rate \citep[e.g.,][]{GirichidisEtAl2016b, SimpsonEtAl2016, KimEtAl2020a,KimEtAl2020b} as well as the effective mixing, in which local inflow and outflow might cancel. The latter point also strongly depends on the gas surface density and the gravitational attraction \citep[e.g.,][]{CreaseyTheunsBower2013,LiBryanOstriker2017}. There is also the complication of a change in geometry if for instance the outflow starts from a very localised point and forms an outflow cone, with effectively increasing area for larger heights \citep[e.g.,][]{MartizziEtAl2016, RecchiaBlasiMorlino2016}. We note that we compare the instantaneous star formation rate directly with the outflow rate at different heights without a more detailed analysis. However, in our setups the typical time for the outflows to reach the measurement height is about $100-300\,\mathrm{Myr}$, which is comparable to the outflow time scales in resolved ISM simulations \citep[][]{HillEtAl2012,KimOstriker2018, KimEtAl2020a} as well as in similar dwarf galaxy setups \citep[e.g.,][]{EmerickBryanMacLow2019} but is short compared to the total simulation time. None the less, we interpret the outflows and the mass loading factor as approximate estimates rather than a precise measure for the direct efficiency of stellar feedback.

We illustrate the formation of outflows in the velocity maps in Fig.~\ref{fig:velz-all-models}, which show the velocity in the $z$ direction in an edge-on view of the galaxy. From left to right we show the models with increasing complexity. From top to bottom we depict the time evolution. Blue regions above the disc and red regions below the disc indicate infall towards the disc, which results from the gravitational attraction. At the earliest time the halo gas indicates infalling material towards the disc. The early star formation burst injects CR energy that drives the outflowing shock front acting against the infall ($0.2-0.5\,\mathrm{Gyr}$). In the grey advection model the inflow is stalled by the generated thermal and CR energy close to the injection region of the CRs. At the end of the simulation a large infall dominates the dynamics. This is expected since the additional energy in CRs can only be transported via advection and mixes with the thermal gas. If diffusion is included the CR energy can diffuse relative to the gas and build up a pressure gradient that drives an outflow beyond the injection sites. The stronger dynamics results in locally driven turbulence and MHD instabilities, which are shown in the perturbations of the velocity. Larger diffusion coefficients imply faster CR transport, which causes an earlier onset of the outflow and initially a larger height of the outflow front. The spectral model shows the fastest moving outflow front despite the lowest star formation rate. After about $\sim0.5\,\mathrm{Gyr}$ the two grey diffusion runs develop a low density outflow cone with high outflow velocities. In the \spec model we observe a filamentary turbulent outflow for about $2\,\mathrm{Gyr}$, after which high-velocity cones of a focussed outflow take the path of least resistance away from the galaxy.

More quantitatively, we compute the amount of outflowing gas as the flux of material through the vertical area at height $\pm 10\,\mathrm{kpc}$ of a galacto-centric cylinder of radius $r=15\,\mathrm{kpc}$. The time evolution of the outflows is shown in Fig.~\ref{fig:outflows-time-evol}. From top to bottom we display the star formation rate, the outflow rate, and the mass loading factor. The delay between the peak in the star formation rate and the onset of outflows reaching a height of $10\,\mathrm{kpc}$ is a few hundred million years, which corresponds to an effective mean outflow speed of $\sim50\,\mathrm{km\,s^{-1}}$ assuming that all star formation occurs in the midplane at $z=0$. The grey model with the large diffusion coefficient (\greydiffhi) reaches the highest outflow rate. The mass loading factor for all simulations quickly reaches values larger than 0.5, indicating that a significant fraction of the injected feedback energy can be converted into outflowing motions. We note the largest mass loading for the spectral model with $\eta=1.5$. At later times, both star formation rates as well as outflow rates decrease in an overall similar manner. Consequently, the mass loading factor does not change perceptibly. The grey model with the smaller diffusion coefficient slowly increases the efficiency of launching an outflow. The models \greydiffhi~and \spec~fluctuate over time with an average mass loading of order unity. The advection only run does not show a noticeable outflow at a height of $10\,\mathrm{kpc}$.

In the spectral model we inject CRs with the same momentum distribution, which means that the initial star burst injects most of the CR energy in all momentum bins. As high-energy CRs diffuse faster we expect TeV CRs to be more extended in comparison to 100-MeV CRs. To illustrate the different distribution in CR energy we show cuts of the CR energy density in three different spectral bins in Fig.~\ref{fig:diffusion-different-energies} at $t=1\,\mathrm{Gyr}$. The upper (middle) panels show edge-on (face-on) views of the CR energy. From left to right we plot the CR momentum bins at 100\,MeV~$c^{-1}$, 1\,GeV~$c^{-1}$, and 1\,TeV~$c^{-1}$. The lower panels show the ratio of the CR energy density of the corresponding bin to the total CR energy density, again for the face-on view. The diffusion coefficient differs by almost two orders of magnitude between the lowest and highest CR momentum bin. The resulting distributions in CR energy reflect this difference. The low energies reside closer to the injection region, i.e. with a large overdensity close to the galactic center. The GeV particles can fill a larger fraction of the domain while still showing a strong overdensity of energy in the centre. Diffusion is very fast for TeV CRs so that they quickly form an almost uniform background energy with only a small enhancement in the centre. The ratio in the bottom panels indicates that the centre is completely dominated by the low-energy CRs. The GeV range dominates the galactic disc in a ring around the centre with a radius of $\sim10\,\mathrm{kpc}$. Beyond that radius, TeV CRs dominate the energy distribution.

The fact that the central region with the highest gas densities coincides with an overdensity of low-energy CRs illustrates the difference between the steady-state CR distribution (see figure~5 of \citealt{EnsslinEtAl2007}) used in grey approaches and the spectrally resolved cooling of CRs. Below a momentum of $\sim1\,\mathrm{GeV}~c^{-1}$ the Coulomb cooling rate increases strongly with a scaling of approximately $p^{-1.9}$ \citepalias{GirichidisEtAl2020} so that low-energy CRs cool fastest in the centre. Because we still find a relative overdensity of low-energy CRs in comparison to GeV CRs, we conclude that the net cooling of the spectrally resolved CRs is less efficient in comparison to the assumed steady-state cooling of grey approaches.

To investigate which CR energies are responsible for the outflow, we pursue two different routes. First, we focus on the outflow front that is launched above and below the centre of the galaxy. In Fig.~\ref{fig:profiles-including-spectral-energies} we show the vertical profiles within a cylinder of radius $5\,\mathrm{kpc}$, centred on the galaxy. From bottom to top we show the density profile, the vertical velocity $\varv_z$, the individual CR energies as well as the individual outward pointing accelerations (colour coded). In the left-hand panels we depict the situation after $200\,\mathrm{Myr}$ while the right-hand panels show a later time ($t=500\,\mathrm{Myr}$). The dip in the velocity profiles towards negative values marks the front between infalling and outflowing/stationary material. This transition coincides with the peak energy and the largest acceleration by CRs with energies between 30 and 100\,GeV~$c^{-1}$. At the early time the energy profiles nearly lie on top of each other and are difficult to distinguish. However, the acceleration peak of the CRs between 30 and 100\,GeV~$c^{-1}$ is clearly visible. At the later time CRs with different momenta separate and we identify two maxima in the acceleration profiles. One that drives the outflow front ($10-100\,$GeV~$c^{-1}$) and one that supports the disc ($\lesssim$GeV~$c^{-1}$).

Second, the kinetic energy-weighted CR momentum is defined via
\begin{equation}
p_\mathrm{peak} = \frac{4\uppi\,\int_{p_\mathrm{min}}^{p_\mathrm{max}}\, p^3 T(p)f(p)\,\mathrm{d}p}{4\uppi\,\int_{p_\mathrm{min}}^{p_\mathrm{max}}\, p^2 T(p)f(p)\,\mathrm{d}p}
\end{equation}
where the minimum and maximum momenta are $p_\mathrm{min}$ and $p_\mathrm{max}$. This momentum indicates the energetically dominant part of the spectrum, which is actively driving the outflows. We compute $p_\mathrm{peak}$ at different heights at the top and bottom of a cylinder of radius $r=15\,\mathrm{kpc}$ centred on the galaxy. Figure~\ref{fig:peak-momentum-diff-heights-time-evol} shows the time evolution of $p_\mathrm{peak}$ at distances ranging from $1$ to $10\,\mathrm{kpc}$ from the midplane (colour coded). We note several systematic trends in the evolution that can be directly connected to the gas dynamics in the galaxy. High-energy CRs diffuse faster to larger heights, which explains the overall monotonic dependency of larger $p_\mathrm{peak}$ at larger heights.

The temporal changes are in line with the outflow behaviour, where we note three different phases. In the beginning ($t\lesssim40$\,Myr) high-energy CRs quickly diffuse to the measurement heights resulting in a large $p_\mathrm{peak}$. The outflow stops infall from the halo early on and reaches a height of $1\,\mathrm{kpc}$ at about $0.1\,\mathrm{Gyr}$. The transported material also contains low-energy CRs, which decreases $p_\mathrm{peak}$ noticeably from initially $\sim200\,\mathrm{GeV}~c^{-1}$ to values around $20\,\mathrm{GeV}~c^{-1}$. The larger the distance, the longer it takes until the outflow front reaches the measurement height. A direct comparison can be done for $h=6\,\mathrm{kpc}$ (dark red line) with the vertical profiles in Fig.~\ref{fig:profiles-including-spectral-energies} at $0.2\,\mathrm{Gyr}$ (left-hand panel). The height at which the infalling gas is stopped by the over-pressurized gas surrounding the forming galactic disc coincides with a noticeable contribution of medium-energy CRs, see dark lines in the top panel of Fig.~\ref{fig:profiles-including-spectral-energies}. Once the infall from the halo is stopped, the remaining CRs start driving a massive outflow into the halo. In this phase progressively lower CR energies become important, which reduces $p_\mathrm{peak}$. Again, a comparison to Fig.~\ref{fig:profiles-including-spectral-energies} illustrates the details. After $0.5\,\mathrm{Gyr}$ (right-hand plot), the measurement height at $10\,\mathrm{kpc}$ shows positive velocities indicating an outflow and a dominating CR momentum contribution between $10$ and $31\,\mathrm{GeV}~c^{-1}$, which corresponds to $p_\mathrm{peak}\sim50-60\,\mathrm{GeV}~c^{-1}$. At late times the injection of fresh CRs reduces further due to the declined star formation rate, which makes low-energy CRs at around $p_\mathrm{peak}\sim10\,\mathrm{GeV}~c^{-1}$ the dominant driver at all heights.

The relative importance of high-energy CRs in comparison to low-energy CRs in determining the dynamics in our setup depends on two aspects. One is the maximum density that can be reached and consequently how much of the CR energy can cool. The other is the environment of the galaxy. If the infalling material is denser and/or the infall faster, then potentially the front of high-energy CRs is not powerful enough to flip the sign from infall to outflow. The impact on the disc that is supported by the low-energy CRs is then larger. If the infall stalls and the central densities do not increase, the low-energy CRs remain important in regulating the central region and thus star formation. Contrary, if the infalling material results in a stronger collapse and much larger central densities, the low-energy CRs, which cannot escape fast enough, can significantly cool and become to weak to regulate the dynamics in the disc.

\section{Spectral evolution}
\label{sec:spectra}

\begin{figure*}
\begin{minipage}{\textwidth}
\includegraphics[width=\textwidth]{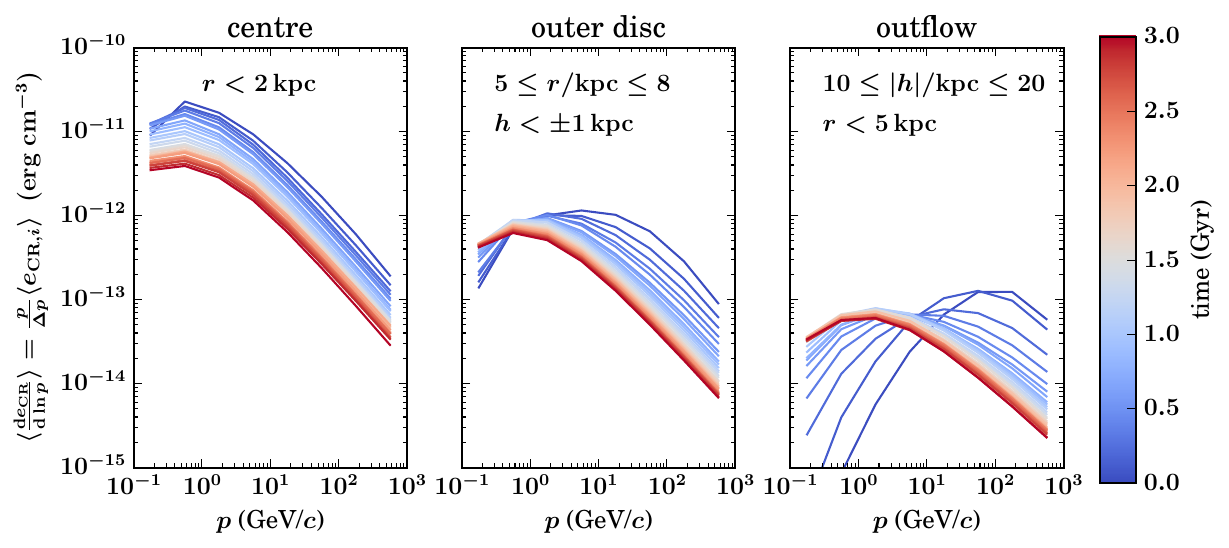}
\includegraphics[width=\textwidth]{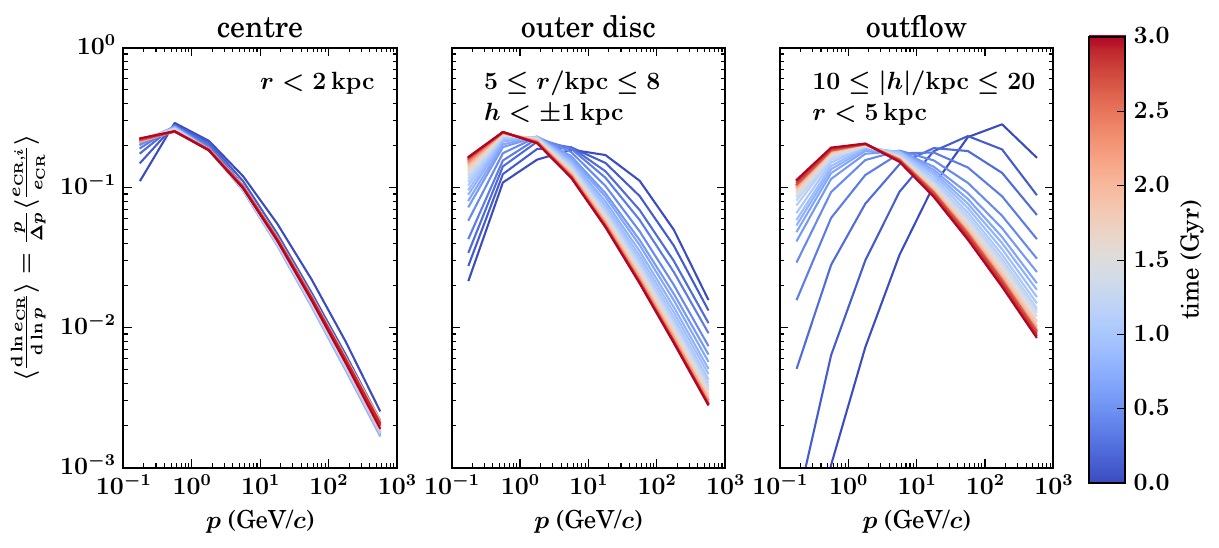}
\caption{Spectral evolution in three different regions (centre, outer disc and outflow) over time (colour coded). The upper panels show the CR momentum spectra. The lower ones show the spectra normalised to the total CR energy in this spatial range. The change in amplitude from centre, to outer disc, to outflow region highlights the time delay in CR transport and different CR cooling processes. The decrease in amplitude in the centre directly reflects the decline in star formation rate over time. The spectra in the centre quickly approach the steady state spectrum. In the outer disc and the outflows, a stable steady state spectrum is reached much later.}
\label{fig:spectra-time-evol-different-regions}
\end{minipage}
\end{figure*}

\begin{figure*}
\begin{minipage}{\textwidth}
\includegraphics[width=\textwidth]{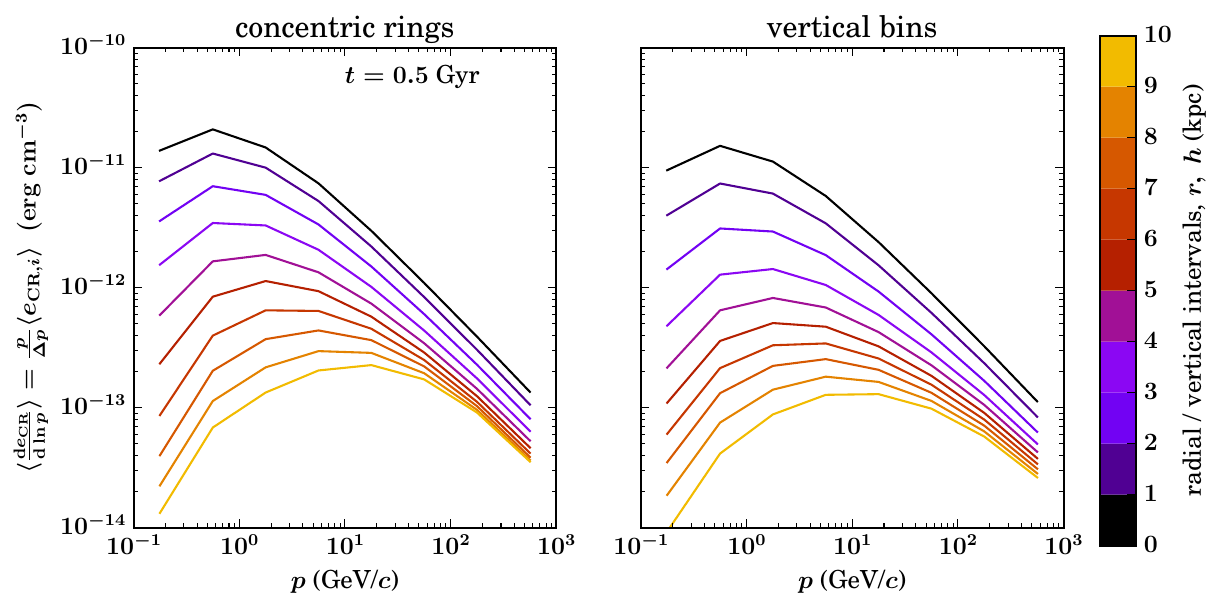}
\includegraphics[width=\textwidth]{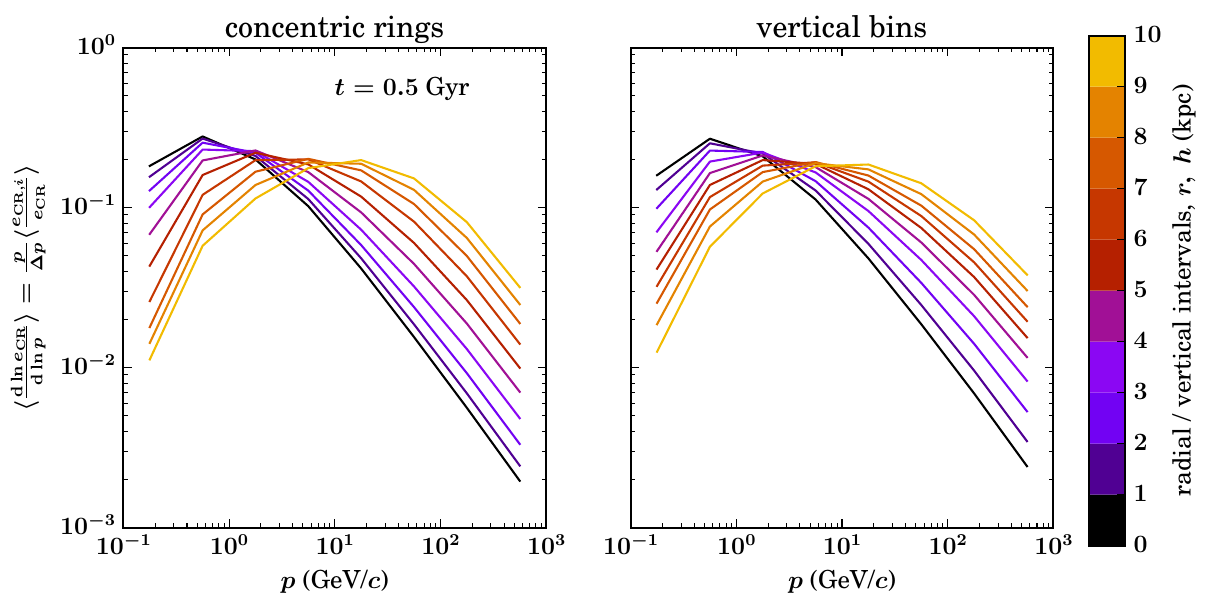}
\caption{Spatial differences of the energy spectra (top) and the normalised CR momentum spectra (bottom) at $t=0.5\,\mathrm{Gyr}$ averaged in concentric rings (left-hand panels) and cylinders at different heights (right-hand panels). Colour coding of the lines labels volume averages over $\Delta r=1\,\mathrm{kpc}$ and $\Delta h=1\,\mathrm{kpc}$. The larger the distance from the centre, the lower is the star formation rate and the smaller the amplitude of the CR energy (top). Faster diffusion of the high-energy CRs causes the spectra at large distances to be increasingly dominated by high CR momenta (bottom). Only close to the centre the spectra show typical shapes of steady state spectra.}
\label{fig:spectra-different-radii-heights}
\end{minipage}
\end{figure*}

Finally, we would like to focus on the details of the spectral evolution over time and in different regions of the galaxy. We inject CRs with the spectral form of a power law $f(p)\propto p^{-4.5}$ over the entire momentum range. The injection regions are coupled to star formation and are thus the densest cells in the simulation. After the injection, there are four competing effects that alter the spectrum, namely cooling, adiabatic processes associated with advection across density gradients, advective mixing of different CR spectra, and momentum-dependent diffusion. Low-energy CRs diffuse slowest, so they remain in the dense injection regions and can cool fastest. We thus expect a low-energy turnover very quickly. Advection and adiabatic compression or expansion do not change the spectral slope. However, the advective transport combined with local mixing determines the effective spectra in different regions. Diffusive transport distributes high-energy CRs faster than the low-energy CRs with a diffusion coefficient that varies by almost two orders of magnitude over the entire spectrum, see Table~\ref{tab:spec-bins}.

Figure~\ref{fig:spectra-time-evol-different-regions} shows spatially averaged CR momentum spectra in different regions of the galaxy over time. The top panels show the energy spectra $\langle \mathrm{d}e_\mathrm{CR} / \mathrm{d}\ln p\rangle$ while we show the spectra normalised to the total CR energy, $\langle \mathrm{d}\ln e_\mathrm{CR} / \mathrm{d}\ln p\rangle$, in the bottom panels. We distinguish three regions of the galaxy, the centre (left-hand panel), the outer disc (middle panel) and the outflow region above the disc (right-hand panel). For the central region we investigate a sphere of radius $2\,\mathrm{kpc}$ around the centre. The outer disc covers a radial extent between $5$ and $8\,\mathrm{kpc}$ and a vertical extent of $\pm1\,\mathrm{kpc}$ from the midplane. For the outflow spectra we chose two cylindrical regions with radius $r<5\,\mathrm{kpc}$ at vertical heights ranging in between $10$ and $20\,\mathrm{kpc}$ above and below the midplane.

The central region shows the fast evolution towards a steady-state spectrum. The normalised energy spectra (bottom) are almost identical over the entire simulation time and show the roll-over at low momenta due to Coulomb cooling and the high-energy spectral index of -5, which is the injection spectrum steepened by diffusion. The amplitude of the total energy spectra in the top panel is directly coupled to the central star formation and CR injection rate. Investigating the outer disc region reveals the delay in advective and diffusive transport of CRs from the central region, where star formation and thus CR injection dominates, out to a radius of 5 to 8\,kpc. While advection of CR energy from the centre to the outer disc would imply an identical spectral shape but with a lower amplitude, diffusive transport allows high-energy CRs to diffuse faster, which leads to larger amplitudes at larger momenta, which is what we observe. The normalised spectra show the evolution towards a similar steady state spectrum as in the centre, but over a longer time scale of approximately $1.5-2\,\mathrm{Gyr}$. A similar behaviour is seen in the outflow region above and below the midplane of the galaxy. The overall amplitude is lower in comparison to the CR energy in the (outer) disc and the spectra are broader with larger high-energy contributions. The low-energy behavior
can be explained by slower Coulomb losses in the low-density circum-galactic medium. The larger contribution of high-energy CRs in particular at early times results from two aspects. First, high-energy CRs drive the initial outflow front (see Figs~\ref{fig:profiles-including-spectral-energies} and \ref{fig:peak-momentum-diff-heights-time-evol}) and second, the orientation of the field in the disc is mainly azimutal and since the diffusion is predominantly along the field lines, the radial diffusion of CRs from the centre to the outer disc is delayed in comparison to the turbulent field in the outflow region.

Finally, we show the radial and vertical differences at a fixed time of $t=0.5\,\mathrm{Gyr}$ in Fig.~\ref{fig:spectra-different-radii-heights}. As in Fig.~\ref{fig:spectra-time-evol-different-regions} we show in the top panels the energy spectra and the bottom panels the normalized counterparts. The spectra in the left-hand panels are spatial averages in concentric rings with intervals of $\Delta r=1\,\mathrm{kpc}$ (colour coded). In the right-hand plots we average over cylinders of radius $r=5\,\mathrm{kpc}$ in vertical bins of width $\Delta h=1\,\mathrm{kpc}$. The spectral evolution confirms the previously described evolution. At larger heights and galacto-centric radius the total CR energy is lower (top panels) and the spectra are dominated by higher-energy CRs, in particular at early times. In the central region we note again that the spectrum is close to the steady state shape.

\section{Discussion}
\label{sec:discussion}

We model CR transport in the advection-diffusion approximation, which is a simplified transport prescription and does not account for the interactions of CRs with self-generated Alfv\'en modes as a result of the streaming instability \citep{KulsrudPearce1969, Wentzel1974,Skilling1971,Skilling1975,Shalaby2021}. Because amplifying resonant Alfv\'en waves requires CR energy, we expect CRs to loose energy at a rate that is comparable to the growth rate of the streaming instablity. However, because diffusive transport is energy conserving, modeling CR transport solely with diffusion precludes modeling the physics of self-generated transport and thus requires the existence of resonant waves to provide the scattering centres, which could be provided through turbulent cascading the energy from larger scales \citep{YanEtAl2004,YanEtAl2008}. Alternatively, the streaming physics could be emulated in a modified diffusion approach \citep{WienerEtAl2013, FarberEtAl2018, BuckEtAl2020, SemenovKravtsovCaprioli2021}, in which CR losses due the excitation of Alfv\'en waves are included and the diffusion speed is limited. Earlier approaches that explicitly modeled CR streaming transport in the one-moment description relied on regularising the streaming speed \citep{SharmaColellaMartin2010}, which implies numerical diffusivity to provide a numerically stable solution. However, this introduces a numerical parameter that is associated with a critical CR gradient length above which the transport is governed by numerical diffusion, which is naturally the case if CRs are injected into a pre-existing population of CRs \citep[see figure 6 of][]{ThomasPfrommer2019}.

Hence, a more fundamental two-moment approach has been introduced that evolves the CR energy and momentum density either by adopting a CR diffusion coefficient that assumes a steady-state wave energy \citep{JiangOh2018,ChanEtAl2019} or additionally by integrating the foward and backward propagating resonant Alfv\'en wave energies which enable a more accuate modeling of the CR diffusion coefficient \citep{ThomasPfrommer2019,ThomasPfrommer2021,ThomasEtAl2021} so that these models self-consistently transfer energy and momentum between the CR, magnetic and thermal energy reservoirs. In addition, these two-moment approaches yield effective CR transport speeds that depend on the efficiency of wave damping and hence the coupling strength of CRs to the background plasma: while strong wave damping implies a fast diffusive CR transport, weak wave damping retains large wave amplitudes that can isotropise the CR population in the Alfv\'en wave frame and implies a transport with the Alfv\'en waves. Whereas this approach is certainly more self-consistent, it requires an accurate estimate of the local degree of ionization in the gas because of the vastly different efficiencies of ion-neutral and non-linear Landau damping of Alfv\'en waves and thus the associated CR transport speeds in order to accurately model self-consistent CR transport. Hence, our CR diffusion approach enables us to control the CR transport speed and to separate the effects of CR self-confinement from CR spectral effects on the morphology of galaxies and the properties of outflows emerging from the galactic discs.

A resolved ISM that includes the various feedback processes from stars and stellar clusters can account for an accurate thermal state of the gas, which is directly connected to the damping processes of Alfv\'en waves and the effective coupling of CRs with the thermal gas. In the vicinity of CR sources such as SN remnants and stellar wind shocks the large CR fluxes drive vigorous plasma instabilities \citep{Bell2004,Shalaby2021} that imply large saturation amplitudes of Alfv\'en waves and hence might reduce the transport speeds to values much lower than predicted by the galactic diffusion coefficient \citep[e.g.,][]{RevilleKirkDuffy2009,TelezhinskyEtAl2012b,AbeysekaraEtAl2020,DiMauroEtAl2020,BrahimiMarcowithPtuskin2020,BustardZweibel2021}. Besides, thermal and radiation feedback in the ISM will alter the magnetic field direction and strength directly and launch local outflows and galactic fountain flows, which will aid in lifting gas out of the disc. Future work that follows a multi-phase gas with {\em resolved gradients} into the dense star-forming cores of molecular clouds are thus required to progress to the next level of spectral CR simulations.

Our spectral CR approach yields steady-state spectra within the simulated time in most regions of the galaxy. The time scale needed to approach this steady state differs strongly between the regions ranging from $\lesssim100\,\mathrm{Myr}$ in the centre to $\sim2\,\mathrm{Gyr}$ at larger distance from the centre. In addition, the CR spectra differ in shape. Steady-state spectra result from a balance of gains and losses, where the gains include local CR injection as well as advective and diffusive transport into the region of interest. The latter two also account for transport losses besides Coulomb and hadronic losses. The resulting steady state spectra are therefore likely to take on a variety of shapes, depending on the local properties of the ISM.

Finally, we like to emphasize that variations of the spectral index of the injection spectrum can quantitatively influence the emerging dynamics via a different pressure, via variations of the energy scaling of diffusion coefficient and via different steady-state spectra. As a result, the effective momentum that drives an outflow could also be altered. We assume a spectral index of $4.5$, which is slightly steeper than the canonical injection index of $4.2$ for strong SNe shocks. We chose this steeper spectrum because the individual SNe are not resolved in our modeling and we account for a small contribution of included diffusive adjustments to the spectrum. However, we do not expect a systematically different outcome for our models for such slight variations in the injection spectrum.

\section{Conclusions}
\label{sec:conclusions}
We implemented our spectrally resolved CR evolution model \citep{GirichidisEtAl2020} into the MHD code \textsc{Arepo}. The spectral evolution of CR protons is followed in each cell with 8 spectral bins from the MeV to the TeV range with accurate CR cooling, adiabatic compression and expansion as well as an improved energy dependent CR diffusion. We optimize the coupling between spectral evolution and spatial diffusion to be a fast and robust method that can be applied in highly dynamical environments. The improved model is then applied to a simplified isolated galaxy with a halo mass of $10^{11}\,\mathrm{M}_\odot$ and an initial gas mass of $1.55\times10^{10}\,\mathrm{M}_\odot$. Four different simulations with different complexities are investigated. Three of them use the grey CR approach with only advection, advetion plus diffusion with a parallel diffusion coefficient of $D = 10^{28}\,\mathrm{cm^{2}\,s^{-1}}$, and advection plus diffusion with $D = 4\times10^{28}\,\mathrm{cm^{2}\,s^{-1}}$. These setups are compared to our new spectral model with $D(p) = 10^{28}\,[p/(1\,\mathrm{GeV}~c^{-1})]^{0.5}\,\mathrm{cm^2\,s^{-1}}$. We find that the dynamical impact of our spectrally resolved model differs in many aspects from the grey CR approaches. Our results can be summarized as follows.

\begin{itemize}
\item Spectrally resolved CRs result in a different morphological appearance in comparison to their distribution in grey diffusion models. The energy dependent transport causes high-energy CRs to diffuse faster through the disc. As a result the outflows are launched from a larger part of the disc rather than primarily from the centre, where most of the star formation occurs.
\item The explicit inclusion of low-energy CRs with smaller diffusion speeds leads to a slower removal of CR pressure close to the star forming regions and the CR injection sites. This reduces the maximum density by an order of magnitude compared to the grey models. As a result, the star formation rate is noticeably lower if CRs are spectrally resolved. In addition, the net CR cooling rate, which scales linearly with the gas density, is lower compared to the cooling in the grey models. Both effects lead to a larger CR pressure in the dense regions that can slow down subsequent star formation and drive outflows from the disc. We note that the relative importance of low-energy CRs in comparison to high energy CRs can depend on the environmental properties of the galaxy and the halo as well as on the detailed conditions in the ISM.
\item We find that high-energy CRs with a momentum of $\sim100\,\mathrm{GeV}~c^{-1}$ can stop the infall of gas from the halo onto the galaxy and launch a first outflow front. Over time, a more massive outflow emerges due to acceleration by the vertical CR pressure gradient, which is provided by CRs with progressively smaller CR energies. The energy-weighted mean momentum that drives the outflow decreases from values of $p_\mathrm{peak}\sim200-600\,\mathrm{GeV}~c^{-1}$ at early times to $p_\mathrm{peak}\sim8-15\,\mathrm{GeV}~c^{-1}$ at late times. We find that $p_\mathrm{peak}$ increases with larger distance from the injection site.
\item The spectral shape differs between the locations in the galaxy and evolves over time. In the central region of the galaxy the CR spectrum quickly approaches a steady-state shape. In the outer disc, a steady-state spectrum only establishes after approximately $2\,\mathrm{Gyr}$. Similar time scales apply for the outflow region above and below the centre of the galaxy. Furthermore, the shape of the steady state spectrum differs between the regions which is due to the local differences in (diffusive) CR supply or injection and local conditions for the losses.
\end{itemize}

In this work, we demonstated for the first time that spectrally resolved CRs have a significant dynamical impact on galaxy formation that differs from the grey CR approaches in terms of star formation rate, the morphology of the galaxy as well as the onset and appearance of galactic outflows. Hence, future accurate models of galaxy formation will require the simulation of CR spectra in space and time.

\section*{Acknowledgements}
The authors thank the anonymous referee for valuable comments that helped improving the paper. PG and CP acknowledge funding from the European Research Council under ERC-CoG grant CRAGSMAN-646955. PG also acknowledges funding from the ERC Synergy Grant ECOGAL (grant 855130).

\section*{Data availability}
The simulation data and data analysis scripts for this study will be shared upon reasonable request to the corresponding author.

%%%%%%%%%%%%%%%%%%%%%%%%%%%%%%%%%%%%%%%%%%%%%%%%%%



\bibliographystyle{mnras}
\bibliography{st_girichidis.bib,st_astro.bib} % if your bibtex file is called example.bib

%\clearpage
%%%%%%%%%%%%%%%%% APPENDICES %%%%%%%%%%%%%%%%%%%%%
\appendix

\section{Derivations of the momentum-dependent adiabatic index}
\label{sec:adiabatic-index-detail}

\subsection{Simplified analytic solution}
We can directly compute the adiabatic index at a momentum $p_0$ by substituting the particle distribution with a Dirac $\delta$ distribution, $f\propto\delta(p-p_0)$. In order to account for the dimensionality and units, we use the general property of the Dirac $\delta$ distribution,
\begin{equation}
\delta(g(x))=\sum_{i} \frac{\delta\left(x-a_{i}\right)}{\left|g^{\prime}\left(a_{i}\right)\right|}
\end{equation}
with a general function $g$, its derivative $g'$ and the zero points $a_i$. In our case, the particle distribution function transforms into
\begin{align}
f(p) &= f_0\delta\left(\frac{p-p_0}{p_0}\right)
= f_0 p_0 \delta(p-p_0).
\label{eq:f_delta}
\end{align}
The normalization $f_0$ is obtained from the CR number density:
\begin{align}
n_\cra = 4\uppi \int_0^\infty p^2 f(p)\dd p = 4\uppi p_0^3 f_0.
\label{eq:ncr}
\end{align}
To proceed we explore how the two parameters of the CR distribution, $p_0$ and $f_0$, transform upon adiabatic changes of the density. Those adiabatic processes conserve the phase space density $d\Gamma=\dd^3 x\,\dd^3 p = \dd (M/\rho)\,\dd^3 p= \mathrm{const.}$, i.e.
\begin{align}
\label{eq:p-rho}
p_0 &= p_1\left(\frac{\rho}{\rho_1}\right)^{1/3},\quad\mbox{and}\\
\label{eq:f-rho}
f_0 &= \frac{n_\cra}{4\uppi  p_0^3 } =\mathrm{const.}
\end{align}
In order to evaluate the adiabatic index at momentum $p_0$ we start with the general ansatz
\begin{align}
\gamma_\cra(p_0) &= \left.\frac{\dd\ln P_\cra}{\dd\ln \rho}\right|_{S}\\
\label{eq:ad-index-split}
&= \frac{\rho}{P_\cra}\,\left[\frac{\partial P_\cra}{\partial p_0}\,\frac{\partial p_0}{\partial \rho} + \frac{\partial P_\cra}{\partial f_0}\,\frac{\partial f_0}{\partial \rho}\right]\\
&= \frac{1}{P_\cra}\,\frac{\partial P_\cra}{\partial p_0}\,\frac{p_0}{3},
\end{align}
where we have used Equations~\eqref{eq:p-rho} and \eqref{eq:f-rho} in the last step. The pressure of our simplified CR distribution (Equation \ref{eq:f_delta}) reads
\begin{align}
P_\cra(p_0) &= \int_0^{\infty}\frac{4\uppi}{3}f_0\frac{p_0\delta(p-p_0)p^4c}{\sqrt{p^2+m_\p^2c^2}}\dd p\\
&= \frac{4\uppi c}{3}f_0\frac{p_0^5}{\sqrt{p_0^2+m_\p^2c^2}}
\end{align}
and the derivative of the pressure with respect to the characteristic momentum reads
\begin{align}
\frac{\partial P_\cra}{\partial p_0} &= \frac{4\uppi c}{3}f_0\left[ \frac{5p_0^4}{\sqrt{p_0^2+m_\p^2c^2}} - \frac{p_0^6}{(p_0^2+m_\p^2c^2)^{3/2}}\right].
\end{align}
Thus, we find for the adiabatic index
\begin{align}
\gamma(p_0) = \frac{5}{3} - \frac{p_0^2}{3(p_0^2+m_\p^2c^2)}.
\end{align}

\subsection{Solution for a piecewise power law distribution}

We can split the total pressure into partial pressures for each spectral bin, where
\begin{align}
  \label{eq:P_cri}
P_\mathrm{cr,i} &= \frac{4\uppi}{3}\, f_{i-1/2}\int_{p_{i-1/2}}^{p_{i+1/2}}  \rkl{\frac{p}{p_{i-1/2}}}^{-\slope_i}\,\frac{p^4c^2}{\sqrt{m_\p^2c^4+p^2c^2}}\dd p\\
& \equiv \frac{4\uppi}{3}\, f_{i-1/2}\,I_{p_{i-1/2}}^{p_{i+1/2}},\\
I_{p_{i-1/2}}^{p_{i+1/2}}&=\frac{ (m_\p c)^4 c}{2}\left(\frac{p_{i-1/2}}{m_\p c}\right)^{q_i}
\left[\B_{x(p)}\left(\frac{5-q_i}{2},\frac{q_i-4}{2}\right)\right]_{p_{i-1/2}}^{p_{i+1/2}}\,.
\end{align}
Here $i$ denotes the bin index of the spectral grid, $\B_{x}\left(a,b\right)$ denotes the incomplete beta function, and $x(p) = p^2/(p^2+m_\p^2c^2)$. The particle distribution function for momentum bin $i$ can be written as
\begin{equation}
f_i(p) = f_{i-1/2}\left(\frac{p}{p_{i-1/2}}\right)^{-q_i} \theta(p-p_{i-1/2})\theta(p_{i+1/2}-p).
\end{equation}
The adiabatic index for momentum bin $i$ is then 
\begin{align}
\gamma_\cri = \frac{\rho}{P_\cri}& \left(\frac{\pp P_\cri}{\pp f_{i-1/2}}\frac{\pp f_{i-1/2}}{\pp \rho}\right.\nonumber\\
&+\frac{\pp P_\cri}{\pp p_{i-1/2}}\frac{\pp p_{i-1/2}}{\pp \rho}  
+\left.\frac{\pp P_\cri}{\pp p_{i+1/2}}\frac{\pp p_{i+1/2}}{\pp \rho}\right).
\end{align}
To proceed we explore how the three parameters of the CR distribution, $p_{i-1/2}$, $p_{i+1/2}$ and $f_{i-1/2}$, transform upon adiabatic changes of the density. Using the conservation of phase space density, we obtain
\begin{equation}
\label{eq:f-rho-dependence}
f_i = f_0\left(\frac{\rho}{\rho_0}\right)^{q_i/3}.
\end{equation}
Using Equation~\eqref{eq:P_cri}, we find for the adiabatic index
\begin{align}
\gamma_\cri = &\frac{q_i}{3} - \frac{1}{3}\frac{p_{i-1/2}^5c}{\sqrt{m_\p^2c^2+p_{i-1/2}^2}}\left(I_{p_{i-1/2}}^{p_{i+1/2}} \right)^{-1}\nonumber\\
& + \frac{1}{3}\frac{p_{i+1/2}^5c}{\sqrt{m_\p^2c^2+p_{i+1/2}^2}}\left(\frac{p_{i+1/2}}{p_{i-1/2}}\right)^{-q_i}\left(I_{p_{i-1/2}}^{p_{i+1/2}} \right)^{-1}.
\end{align}

As a consistency check we derive the low and high momentum limit of the equation. For the low momenta we approximate
\begin{align}
\sqrt{p^2c^2+m_\p^2c^4} &\rightarrow m_\p c^2,\\
I_{p_{i-1/2}}^{p_{i+1/2}} &\rightarrow \int_{p_{i-1/2}}^{p_{i+1/2}} \left(\frac{p}{p_{i-1/2}}\right)^{-q_i}\frac{p^4}{m_\p}\dd p.
\end{align}
The adiabatic index is then
\begin{align}
\gamma_\cri &= \frac{q_i}{3}-\frac{1}{3m_\p}\left[p_{i-1/2}^5 - p_{i+1/2}^5\left(\frac{p_{i+1/2}}{p_{i-1/2}}\right)^{-q_i}\right]\left(I_{p_{i-1/2}}^{p_{i+1/2}}\right)^{-1}=\frac{5}{3},
\end{align}
which is the expected result for a non-relativistic gas. For the relativistic limit we use
\begin{align}
\sqrt{p^2c^2+m_\p^2c^4} &\rightarrow pc,\\
I_{p_{i-1/2}}^{p_{i+1/2}} &\rightarrow \int_{p_{i-1/2}}^{p_{i+1/2}} \left(\frac{p}{p_{i-1/2}}\right)^{-q_i}{p^3c}\,\dd p,
\end{align}
and the resulting adiabatic index reads
\begin{align}
\gamma_\cri &= \frac{q_i}{3}-\frac{c}{3}\left[p_{i-1/2}^4 - p_{i+1/2}^4\left(\frac{p_{i+1/2}}{p_{i-1/2}}\right)^{-q_i}\right]\left(I_{p_{i-1/2}}^{p_{i+1/2}}\right)^{-1}=\frac{4}{3},
\end{align}
as expected for a relativistic gas.

\section{Numerical tests}
\label{sec:app-tests}

\subsection{Modified sound waves}

\begin{figure}
\includegraphics[width=8cm]{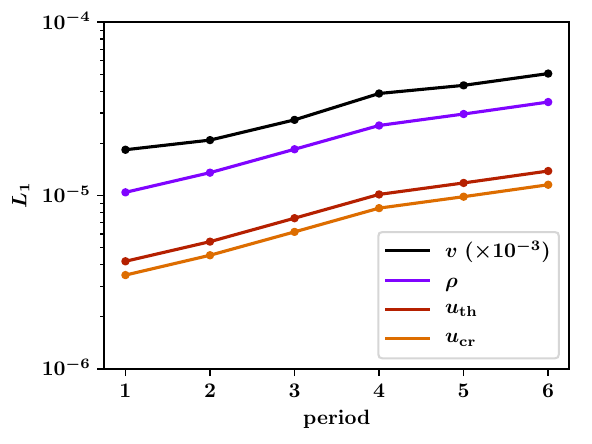}
\caption{$L_1$ error norm of the modified sound wave as a function of the wave periods. Thermal and CR energy as well as the density show errors of the order of $10^{-5}$. The $L_1$ norm of the velocity is significantly larger with values of $\sim10^{-2}$.}
\label{fig:test-wave-error}
\end{figure}

We test the evolution of a modified sound wave through a two component medium with thermal gas and CRs \citep{RaseraChandran2008,YangEtAl2012}. We set up a one-dimensional box with a thermal and CR background pressure, $P_{\mathrm{th},0}=1$ and $P_{\cra,0}=1$, respectively, in a homogeneous medium with density $\rho_0=1$ at rest, $\varv_0=0$. The adiabatic wave speed is given by
\begin{equation}
c_\mathrm{s} = \sqrt{\frac{\gamma_\mathrm{th}P_{\mathrm{th},0} + \gamma_\cra P_{\cra,0}}{\rho_0}}.
\end{equation}
We then add a perturbation wave with a velocity amplitude $\delta \varv=10^{-3}$ and a wavelength of half the box size. Eigenfunctions of the system are triggered by using
\begin{align}
\frac{\delta \rho}{\rho_{0}} &=\frac{\delta \varv}{c_\mathrm{s}}, \\
\frac{\delta P_\mathrm{th}}{P_{\mathrm{th},0}} &=\gamma_\mathrm{th} \frac{\delta \varv}{c_\mathrm{s}}, \\
\frac{\delta P_{\cra}}{P_{\cra,0}} &=\gamma_{\cra} \frac{\delta \varv}{c_\mathrm{s}}.
\end{align}
We show the $L_1$ norm of the error for several periods of the wave in Fig.~\ref{fig:test-wave-error}. The errors for the density, the gas and CR energy are of the order of $10^{-5}$ over six wave periods. The error in the velocity is significantly larger with an accuracy of only $10^{-2}$.

\subsection{Energy dependent diffusion}

\begin{figure}
\includegraphics[width=8cm]{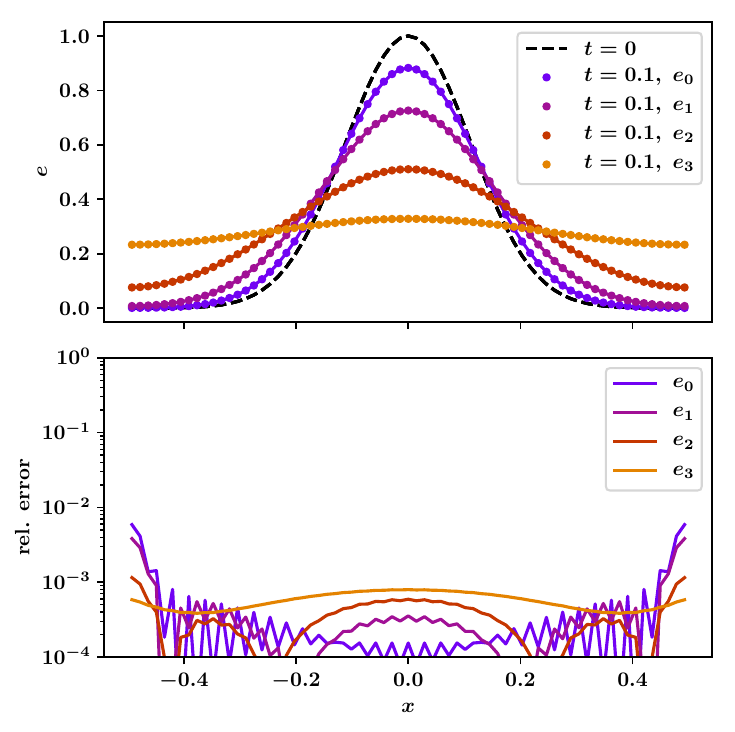}
\caption{Test of the energy dependent diffusion. The top panel shows the spatial energy distribution at $t=0$, which is the same for all CR momentum bins as well as the distribution at $t=0.1$ for all bins. The relative error in the bottom panel is below the percent level for all momentum bins and all spatial coordinates.}
\label{fig:diff-2D-error}
\end{figure}

We modify the diffusion problem to only follow the energy density in each spectral bin rather than both number and energy density, see Section~\ref{sec:numerics}. We test the diffusion solver by using a Gaussian distribution, which takes the time-dependent form
\begin{equation}
  F(x, t) = A \sqrt{\frac{\sigma^2}{\sigma^2+4 D t}}\exp\rkl{-\frac{x^2}{\sigma^2+4 D t}}.
  \label{eq:diff}
\end{equation}
In order to avoid boundary effects for the comparison of our numerical to the analytic solution, we superpose three Gaussian distributions, where the main distribution is centred at the centre of the domain and the other two Gaussians on either side are shifted by a box length
\begin{equation}
e_\mathrm{CR}(x, t) = F(x+L_\mathrm{box}, t) + F(x, t) + F(x-L_\mathrm{box}, t).
\end{equation}
We set up four different spectral bins with an initial amplitude of $A=1$ for all spectral bins. We use the four diffusion coefficients $D = 0.018, 0.056, 0.18, 0.56$ and evolve the system for a time of 0.1. The accuracy of the energy dependent diffusion is illustrated in Fig.~\ref{fig:diff-2D-error}, where we show the spatial energy distribution in the top panel and the relative error to the analytic solution of Equation~\eqref{eq:diff} in the bottom panel, which is below the per cent level across the spatial domain for all diffusion coefficients.

\subsection{Freely cooling and steady state spectrum}

\begin{figure}
\includegraphics[width=8cm]{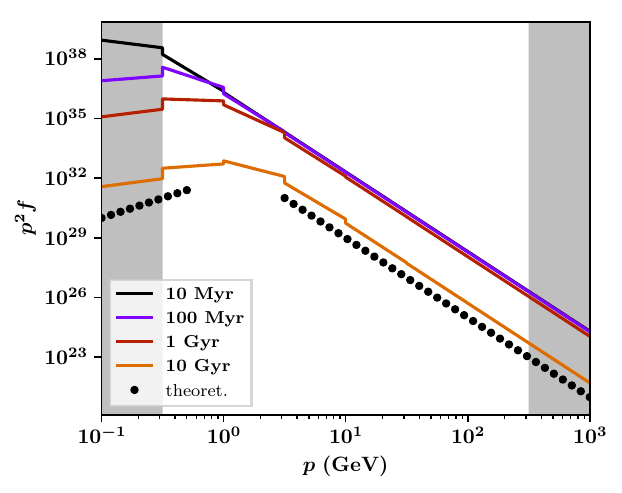}
\caption{Freely cooling CRs. We repeat the free cooling test coupled to the quantities in Arepo with a gas number density of $n=0.01\,\mathrm{cm}^{-3}$. The low and high-momentum slopes are recovered very well.}
\label{fig:test-free-cooling}
\end{figure}

\begin{figure}
\includegraphics[width=8cm]{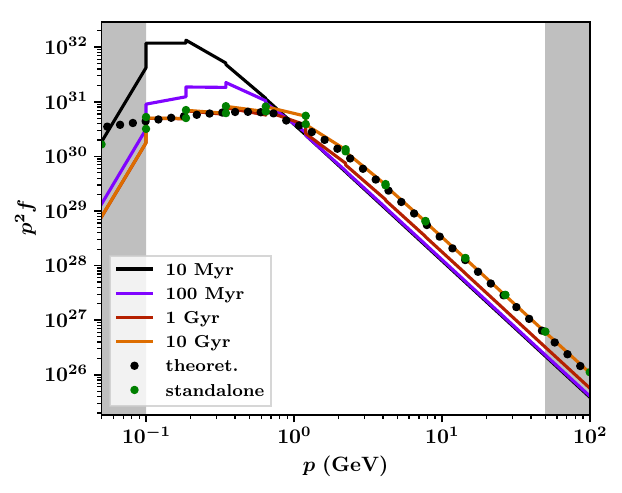}
\caption{Steady-state spectrum. We repeat the steady-state test coupled to the quantities in Arepo with a gas number density of $n=0.01\,\mathrm{cm}^{-3}$. Shown are different times as well as the theoretical and numerical solution from the stand-alone code.}
\label{fig:test-steady-state}
\end{figure}

We repeat the one-zone freely cooling and steady-state test of \citetalias{GirichidisEtAl2020} within Arepo including the coupling of the cell quantities to the spectral solver. The results for the freely cooling are shown in Fig.~\ref{fig:test-free-cooling} with the spectral evolution over time. There is no analytic solution for the amplitude as a function of time. We therefore show the approximate slopes for the high and low-momentum part of the freely cooling spectrum, towards which the numerical solution converges. For the steady-state spectrum we derive the analytic solution in \citetalias{GirichidisEtAl2020} and show the numerical results in Fig.~\ref{fig:test-steady-state}. As expected, the numerical results converge towards the analytic estimate as in the stand-alone code.

% Don't change these lines
\bsp	% typesetting comment
\label{lastpage}

\clearpage
\thispagestyle{empty}
\newgeometry{left=0cm,right=0cm,top=0cm,bottom=0cm}
\includegraphics{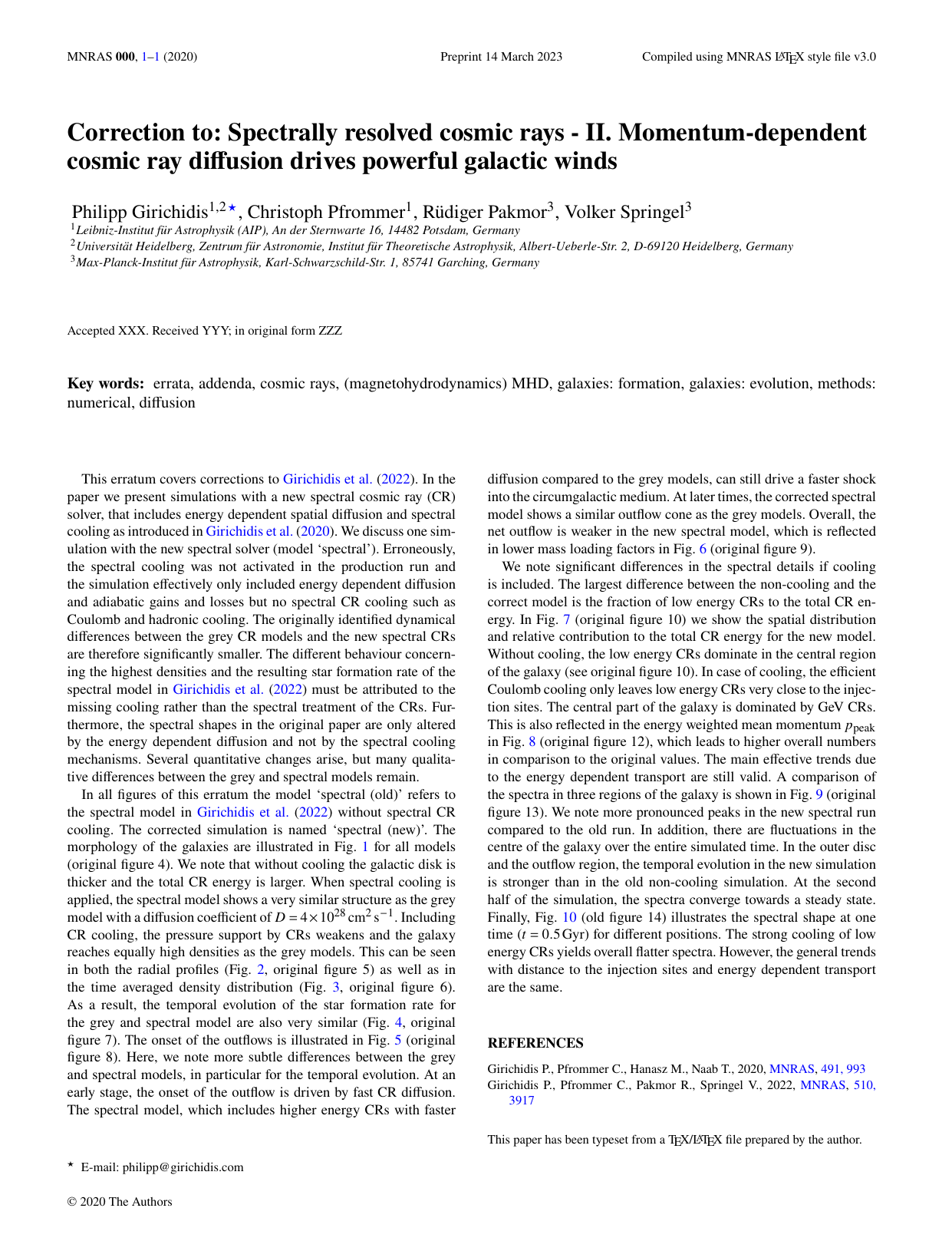}
\newpage
\thispagestyle{empty}
\newgeometry{left=0cm,right=0cm,top=0cm,bottom=0cm}
\includegraphics{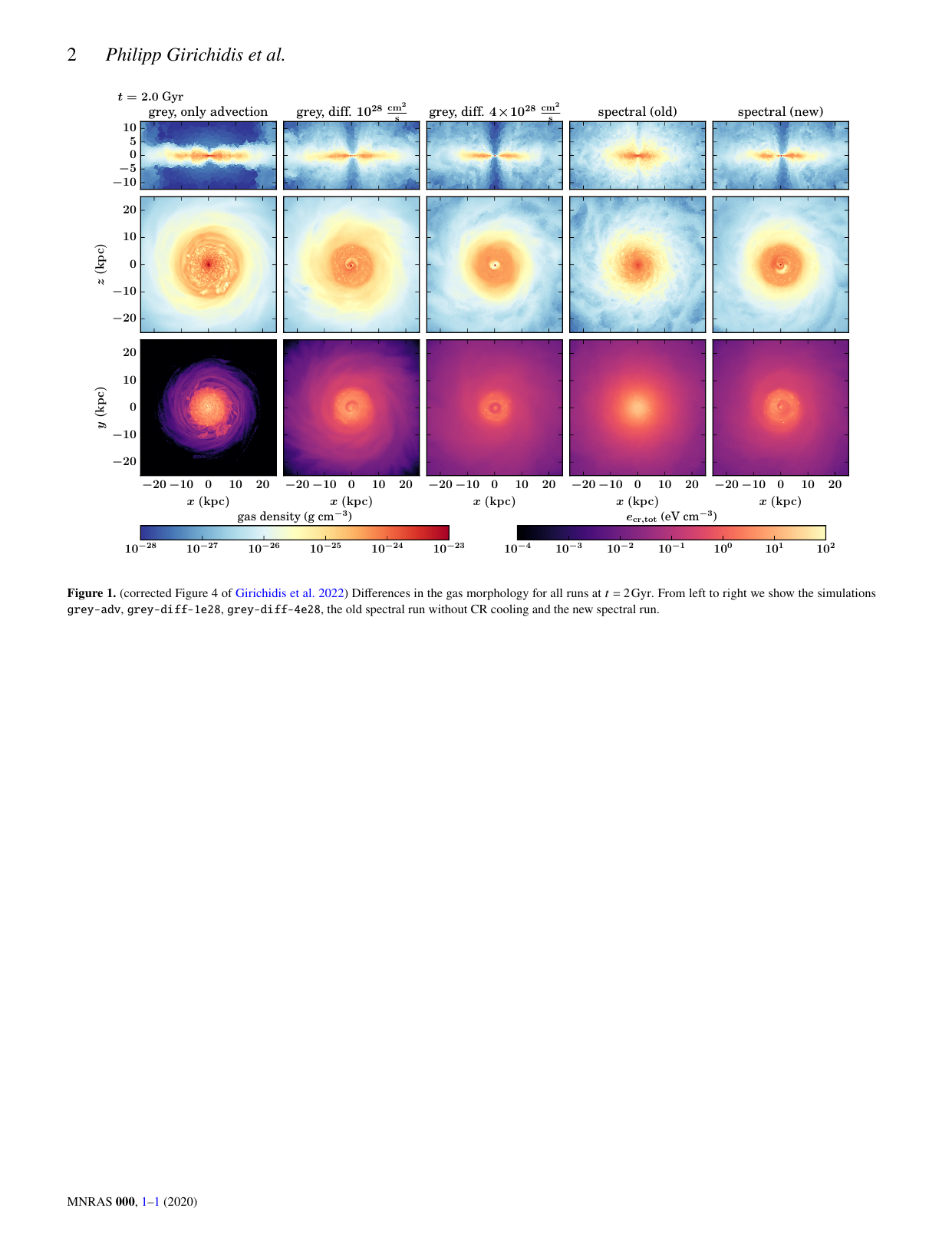}
\newpage
\thispagestyle{empty}
\newgeometry{left=0cm,right=0cm,top=0cm,bottom=0cm}
\includegraphics{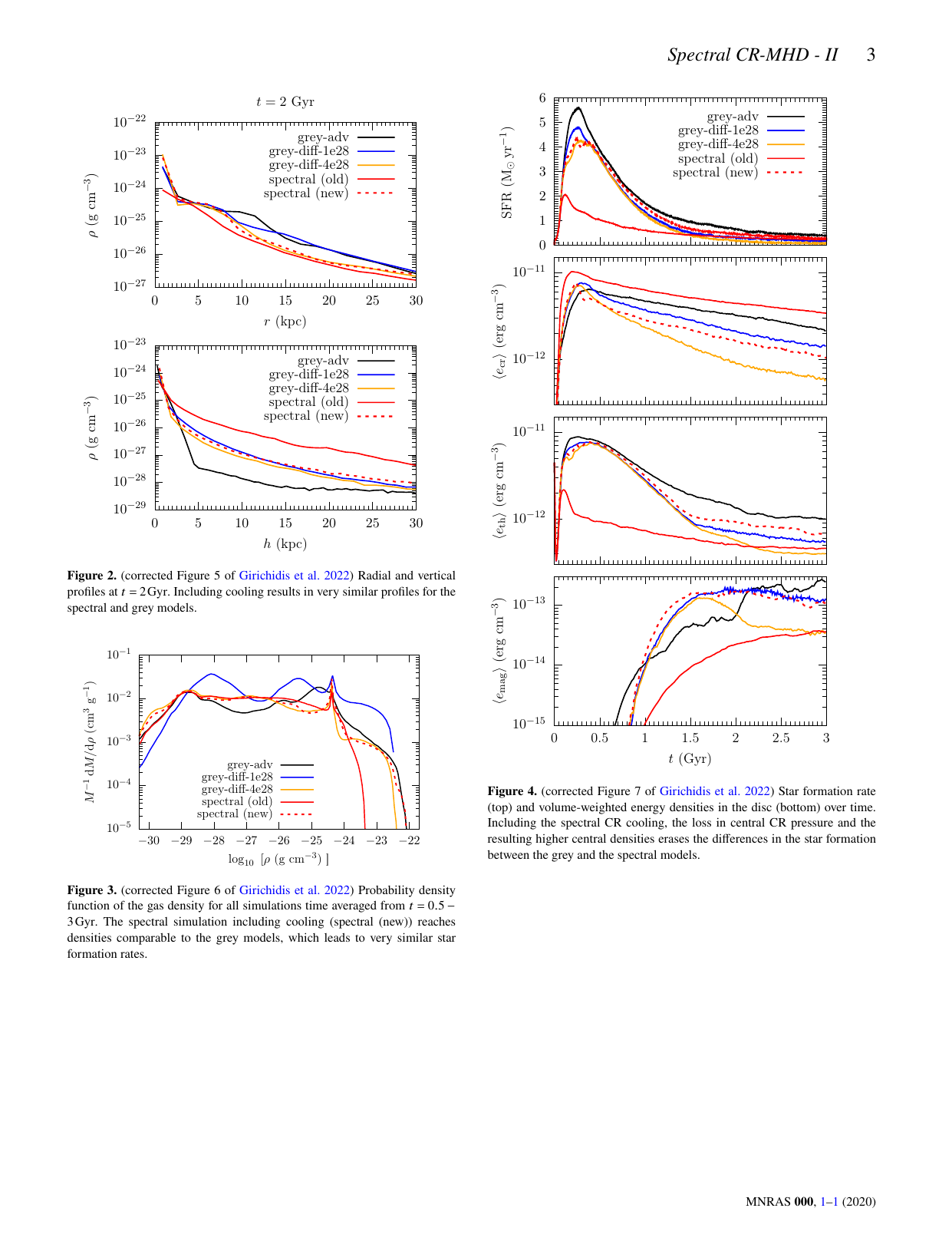}
\newpage
\thispagestyle{empty}
\newgeometry{left=0cm,right=0cm,top=0cm,bottom=0cm}
\includegraphics{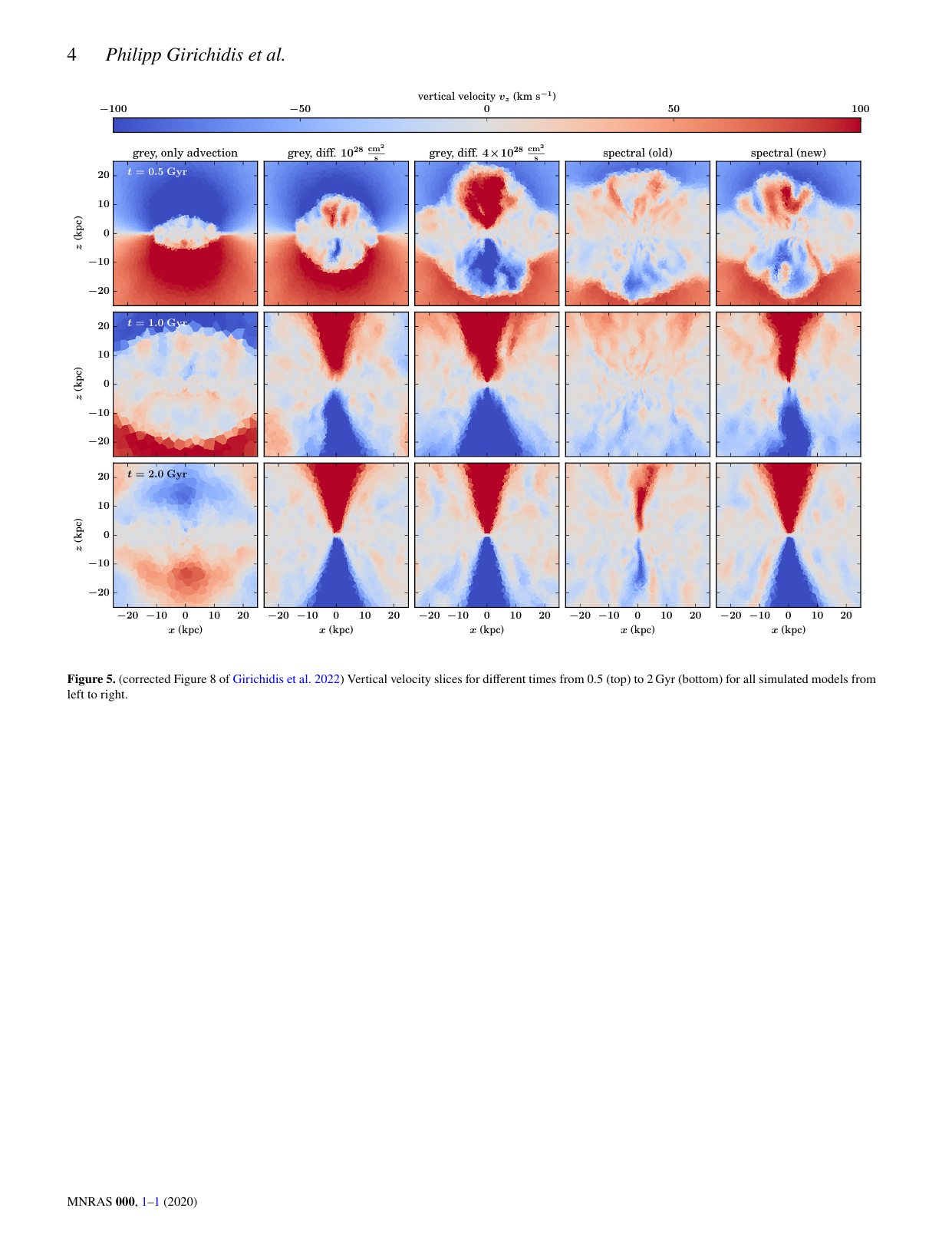}
\newpage
\thispagestyle{empty}
\newgeometry{left=0cm,right=0cm,top=0cm,bottom=0cm}
\includegraphics{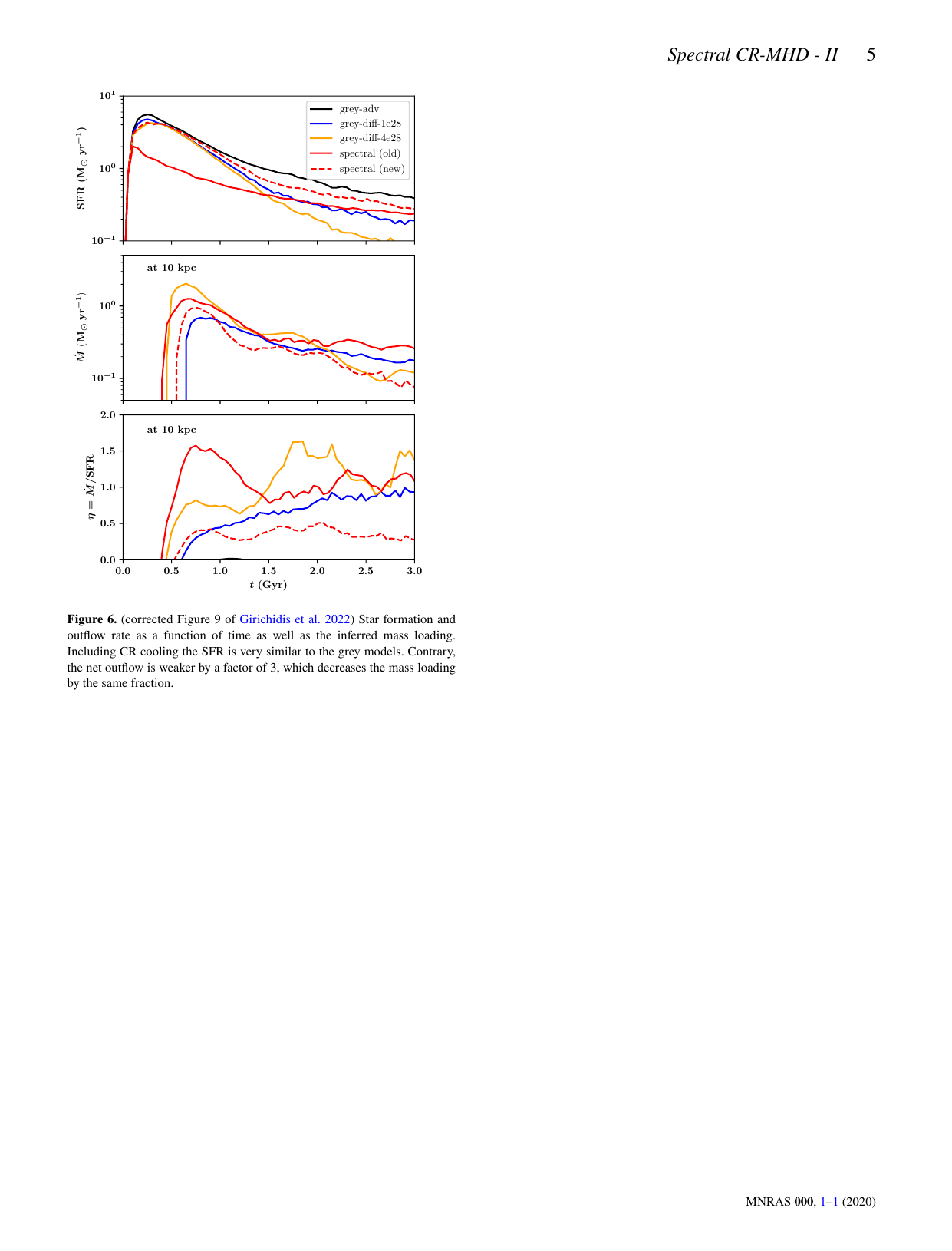}
\newpage
\thispagestyle{empty}
\newgeometry{left=0cm,right=0cm,top=0cm,bottom=0cm}
\includegraphics{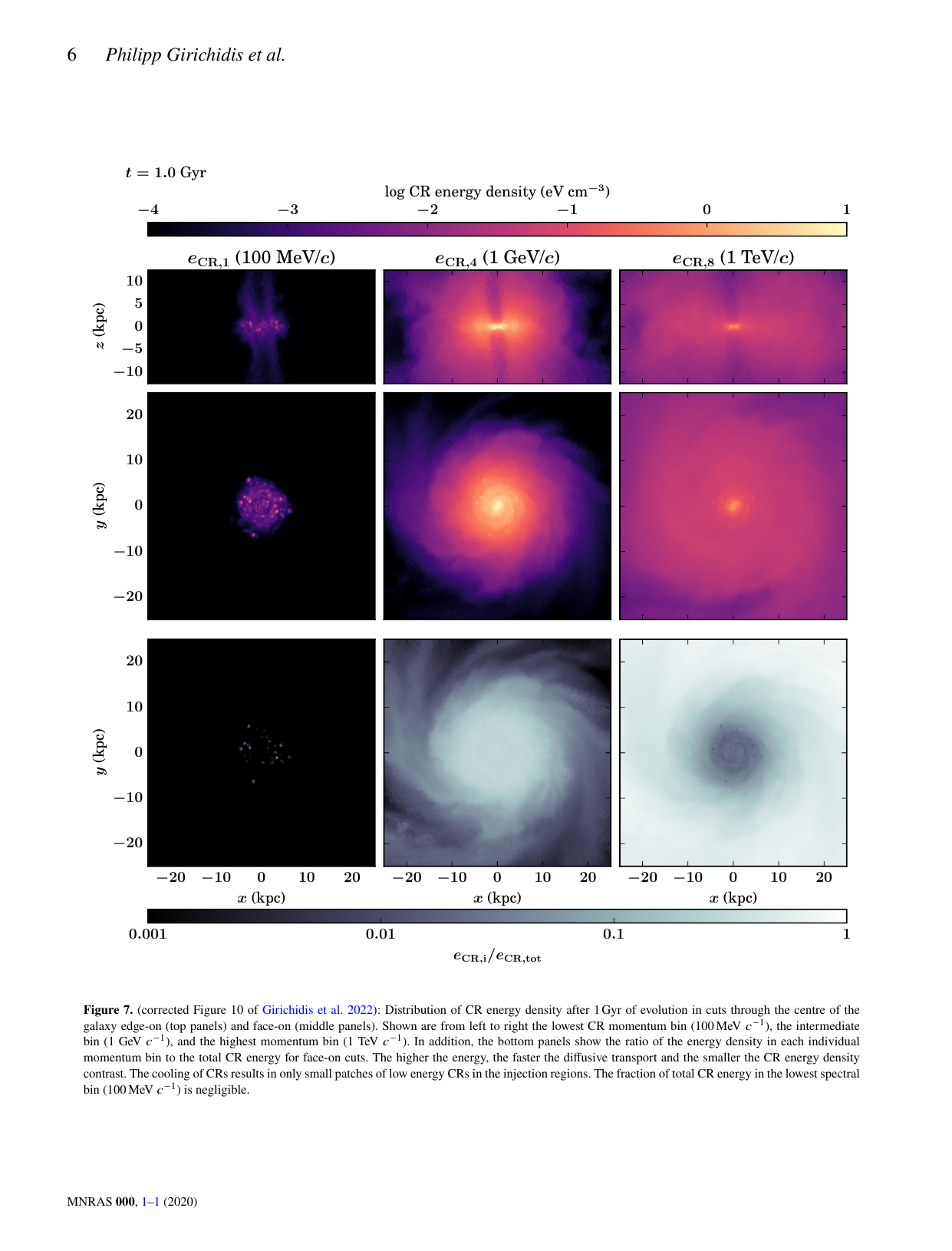}
\newpage
\thispagestyle{empty}
\newgeometry{left=0cm,right=0cm,top=0cm,bottom=0cm}
\includegraphics{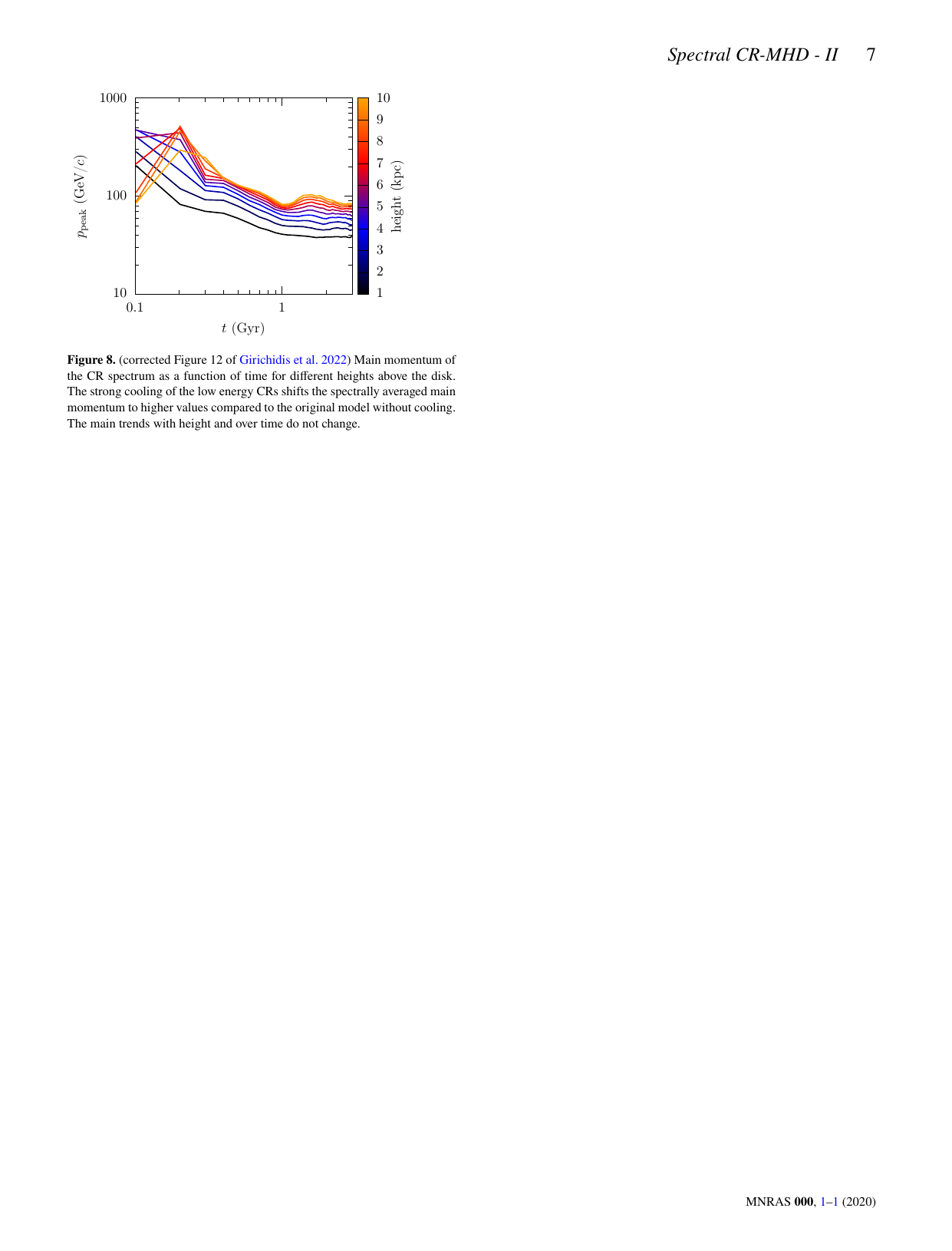}
\newpage
\thispagestyle{empty}
\newgeometry{left=0cm,right=0cm,top=0cm,bottom=0cm}
\includegraphics{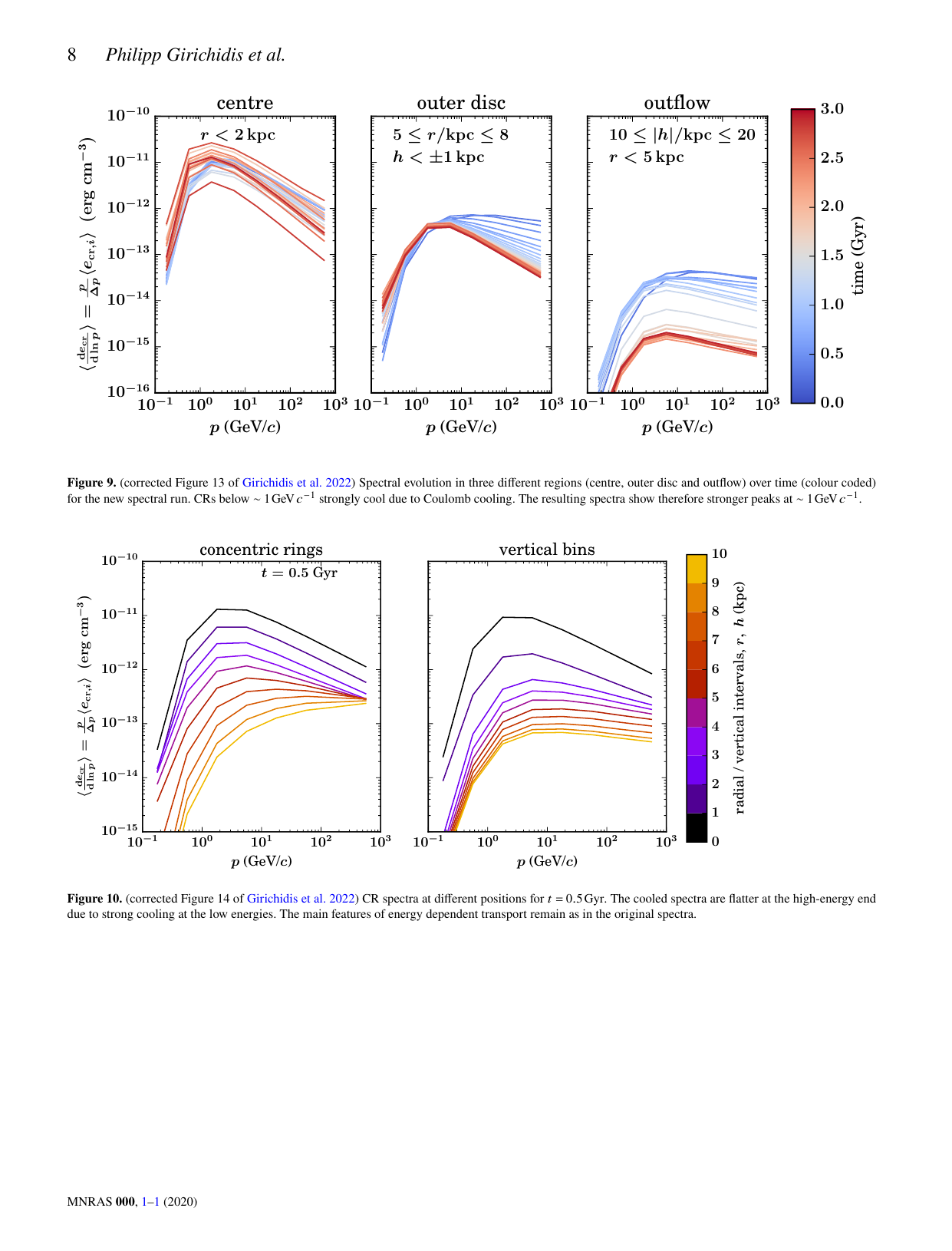}
\end{document}